\documentclass{aa}  

\usepackage{graphicx}
\usepackage{txfonts}
\usepackage{hyperref}
\begin{document}

   \title{Variability and stellar pulsation incidence in Am/Fm stars using TESS and \textit{Gaia} data}

   \subtitle{}

   \author{O. Dürfeldt-Pedros
          \inst{1}
          \and
          V. Antoci\inst{1}
          \and 
          B. Smalley\inst{2}
           \and 
          S. Murphy\inst{3}
           \and 
          N. Posilek\inst{4} 
           \and 
          E. Niemczura\inst{4} 
          }
   \institute{DTU Space, Technical University of Denmark, Elektrovej 327, Kgs. Lyngby, 2800, Denmark\\
              \email{durfeldtpedros.o@gmail.com}   
         \and
             Astrophysics Group, Keele University, Staffordshire, ST5 5BG, United Kingdom 
        \and
            Centre for Astrophysics, University of Southern Queensland, Toowoomba, Qld 4350, Australia 
        \and
            Instytut Astronomiczny, Uniwersytet Wroc{\l}awski, Kopernika 11, 51-622
Wroc{\l}aw, Poland
             }

   \date{Received December 22, 2023; accepted August 19, 2024}

% \abstract{}{}{}{}{} 
% 5 {} token are mandatory
 
  \abstract
   {}
   {We aim to study chemically peculiar Am and Fm stars, distinguished by their unique abundance patterns, which are crucial for studying mixing processes in intermediate-mass stars. These stars provide a window into the atomic diffusion in their stellar envelopes, the evolution-dependent changes in mixing, and the resulting effects on pulsation mechanisms. }
   {This study examines the pulsation characteristics of the Am/Fm star group. Our analysis encompasses 1276 stars (available as catalogues on GitHub), utilising data from TESS and \textit{Gaia} and focusing on stars from the Renson catalogue. }
   { In our sample, 51\% (649 stars) display no variability, thus categorised as constant stars. Among the remaining, 25\% (318 stars) are pulsating Am/Fm and $\rho$\,Puppis stars, including 20\% (261 stars) that are exclusively Am/Fm stars. Additionally, 17\% (210 stars) show variability indicative of binarity and/or rotational modulation and 7\% (93 stars) are eclipsing binaries. Of the pulsating stars, 10\% (32 stars) are $\gamma$ Doradus type, 54\% (172 stars) $\delta$ Scuti type, and 36\% (114 stars) are hybrids, underlining a diverse pulsational behaviour of Am/Fm stars.}
   {Our findings indicate that pulsating stars predominantly occupy positions near the red edge of the classical instability strip, allowing us to ascertain the incidence of pulsations in this stellar population.}

   \keywords{ Stars: chemically peculiar --
                Stars: oscillations --
                binaries: general
               }

   \maketitle
%
%-------------------------------------------------------------------

\section{Introduction}
Intermediate-mass A and F stars have masses 1.4--2.6~$M_\odot$ and can be found on the main sequence, pre main sequence and post main sequence. These are located at the lower part of the classical instability strip, a region of high pulsational activity. The $\delta$ Scuti stars oscillate in p-modes, driven by the $\kappa$-mechanism \citep{Cox1963, Turcotte2000} originating in the He\,II ionisation layer as well as turbulent pressure found in the H/He\,I ionisation layer \citep{Antoci2014, Antoci2019}. Recently, \cite{Murphy2020} suggested that the "edge-bump" mechanism, as described by \cite{Stellingwerf1979}, may also contribute to the excitation of pulsations in Am/Fm stars\footnote{This is caused by a discontinuity in the bound-free opacity at the edge of the hydrogen ionisation zone.}. The frequency range of these oscillations is most commonly defined as 4--80 d$^{-1}$, \citep[e.g.,][]{Kurtz}. 
The $\gamma$ Doradus stars exhibit low-frequency g-mode pulsations that are driven by convective flux modulation at the base of the surface convective zone \citep{Guzik2000, Dupret2005}. Such modes are mostly found at low frequencies below 5~d$^{-1}$. Hybrid stars, pulsating both in p and g-modes, have also been observed in great numbers \citep[e.g.,][]{Griga2010, Uytterhoeven2011, Balona2011}. They are especially interesting targets as their pulsations allow us to study the near-core environment and the outer envelope in detail. Finally, the roAp stars are a specific subclass of chemically peculiar pulsating stars with very strong magnetic fields oscillating in high radial order p-modes \citep{Kurtz1982}. 

A and F stars have convective cores and radiative envelopes with a thin convective outer layer. As the spectral type transitions to early A stars, the convective layer becomes shallower. In the case of slow rotation, there is therefore little mixing occurring in the star and atomic diffusion \citep{Michaud1970, Michaud1982} can take place, causing gravitational settling of He and Ca and radiative levitation of metals with rich spectra in the UV \citep[e.g.][]{Abt}. This process leads to the formation of chemical peculiarities in the observed atmospheres of stars as can be seen in Am and Fm stars, also known as metallic A and F stars \citep{Kurtz1989}. They have over-abundances of metals such as Zn, Sr, Y, Zr and Ba and clear under-abundances of Ca and Sc \citep[e.g.][]{Joshi}. A slow rotation rate below 120~km\,s$^{-1}$ is required for atomic diffusion and has been confirmed in many Am/Fm stars \citep{Abt, Catanzaro2015}. Such rotation rates  can either be intrinsic or induced through tidal breaking by a binary companion. The latter is commonly observed, with 60--70\% of Am/Fm stars estimated to be found in binary systems \citep{Debernardi2000, Smalley2014}. Despite the widespread binary nature, there remains a significant number of Am/Fm stars for which a companion is not known.

The HgMn stars are late B-type stars having strong Hg\,II and Mn\,II absorption lines. Their spectra compare to those of hot Am/Fm stars with early A-type hydrogen lines. They are also primarily found in binary systems, with an incidence of 67\%, and can be perceived as the counterpart of Am/Fm stars in earlier-type stars \citep[e.g.][]{SSC}. 
Chemically peculiar stars are widely present among A-type stars, with 30\% having some form of peculiarity \citep{SSC}. Since the main objective of this paper is the study of the atomic diffusion and stellar pulsations, we focus in this section on a short description of the related peculiarity groups. The $\rho$ Puppis stars are evolved, mostly mid F-type. They have similar chemical abundances as observed in Am/Fm stars \citep{GandG1989} but are more luminous. They could therefore be evolved Am/Fm stars \citep{Kurtz1976} observed in a transition phase where the convective layers deepen and mixing becomes more efficient. Some studies suggest $\rho$ Puppis stars might develop the peculiarities through accretion from a companion, much like Barium stars \citep{McGahee}. However, extensive radial velocity measurements of the prototype $\rho$ Puppis show no indication of a companion \citep{Antoci2013}. 
Other peculiar A and F type stars are Ap and $\lambda$ Bootis stars. Ap stars have strong magnetic fields and are slow rotators. They experience atomic diffusion much like Am/Fm stars \citep{Michaud1970}, with the chemical peculiarities forming along the magnetic field lines and producing spots on the stellar surface \citep{Michaud1981}. Over-abundances of Cr, Sr and rare-Earth elements are typically found \citep[e.g.][]{Romanovskaya}. They are known to pulsate in the form of roAp stars. 
$\lambda$ Bootis stars are metal-weak A type stars \citep{LambdaBoo}, supposed to be acquired  through accretion, with different scenarios having been proposed \citep[e.g.,][and references therein]{MurphyLambdaBoo}.

The Renson catalogue of chemically peculiar stars \citep{Renson} compiles information about known Ap, HgMn and Am/Fm stars. It contains over 8000 objects gathered from literature, with a little more than half being Am/Fm candidates, thereby offering an extensive sample of chemically peculiar stars for ensemble studies but requiring additional verification as they may not all be bona fide detections.

Since atomic diffusion depletes He from the outer stellar envelope, it is expected to (partially)  impede the $\kappa$-mechanism in the He\,II ionisation layer driving most of the $\delta$ Scuti pulsations \citep{Turcotte2000}. For this reason, stellar oscillations were not expected to take place in Am/Fm stars, but have since been robustly detected \citep[e.g.,][]{Balona2011, Smalley2017, Antoci2014, Antoci2019}. Am/Fm stars thus provide asteroseismology with another interesting population of stars to analyse, since these remain poorly studied and challenge the current understanding of pulsation mechanisms. In recent years, turbulent pressure has been proposed to explain the observed pulsations in Am/Fm stars in some parts of the instability strip \citep{Antoci2014, Antoci2019, Smalley2017}. The launch and operation of the Transiting Exoplanet Survey Satellite (TESS) mission \citep{Ricker2015} has provided the asteroseismic community with high quality data to further our understanding of stellar oscillations and stellar structure. 
TESS data has proven crucial due to its vast, homogeneous sample, that opens up many opportunities to study intermediate-mass stars \citep[e.g.,][]{Cunha2019, Antoci2019}.

Precise knowledge of astrophysical and astrometric parameters is fundamental for any research in stellar astronomy, and the European Space Agency (ESA)'s \textit{Gaia} mission \citep{Gaia} has revolutionised the field since its launch. With the recent publishing of \textit{Gaia} Data Release 3 (DR3) \citep{GaiaDR3}, it provides unprecedented accuracy in parallax determinations as well as a wealth of additional data products to be used by the scientific community. 

In this paper, we investigate the pulsational characteristics of the Am/Fm stellar population in a survey of 1276 stars using TESS and \textit{Gaia} data based on stars from the Renson catalogue. We present key statistics from our study and assess whether observational data supports the assumption that turbulent pressure drives oscillations in Am/Fm stars.

%--------------------------------------------------------------------
\section{Data}
In this section, we describe the data used to analyse our sample of Am/Fm stars. Our main focus is to obtain a statistically significant population of stars with minimal sources of bias introduced by our filtering steps, since they can affect our statistical analysis and conclusions regarding the nature of stellar oscillations in Am/Fm stars.
 
\subsection{TESS photometry}
TESS has been obtaining photometric data of stars with the aim of exoplanet detection for over 5 years. The sky is split up into sectors, each being observed for 27 days at a time. The observing campaign started in the southern hemisphere, with TESS revisiting each sector to obtain longer time series for these targets. The observing pattern leads to two regions of the sky being observed for long periods of times, the so-called continuous viewing zones (CVZ) in the southern and northen hemispheres, providing long-uninterrupted light curves ideal for studying stellar oscillations. The data are available through the Mikulski Archive for Space Telescopes (MAST)\footnote{\url{https://archive.stsci.edu/}}, from which 2-min cadence single aperture photometry (SAP) light curves were downloaded for our sample of stars. We used Lightkurve, a Python package for \textit{Kepler} and TESS data analysis \citep{Lightkurve}, to perform an automated query and download of TESS data over a total of 58 sectors. The data are centred around a mean flux of 0 and scaled to the unit of parts per thousand (ppt) for each sector before being stitched together to create longer time series. From these it is possible to compute the amplitude spectrum for each star using a Lomb-Scargle periodogram \citep{Lomb, Scarlge}, which will then be used for frequency analysis and mode identification. 

The number of observed sectors varies significantly within our target list, with only 1 to 2 sectors being available for 70\% of the stars (see Table \ref{tab:AmStars}, full data is given in the supplementary material). The detection and statistical analysis of $\gamma$ Doradus pulsations will thus not be feasible, since g-modes require longer time series to be resolved properly. We provide example light curves and Fourier spectra in Fig. \ref{Fig:Variability} (see Section 4).

\begin{figure*}
   \centering
   \sidecaption
   \includegraphics[width=12cm]{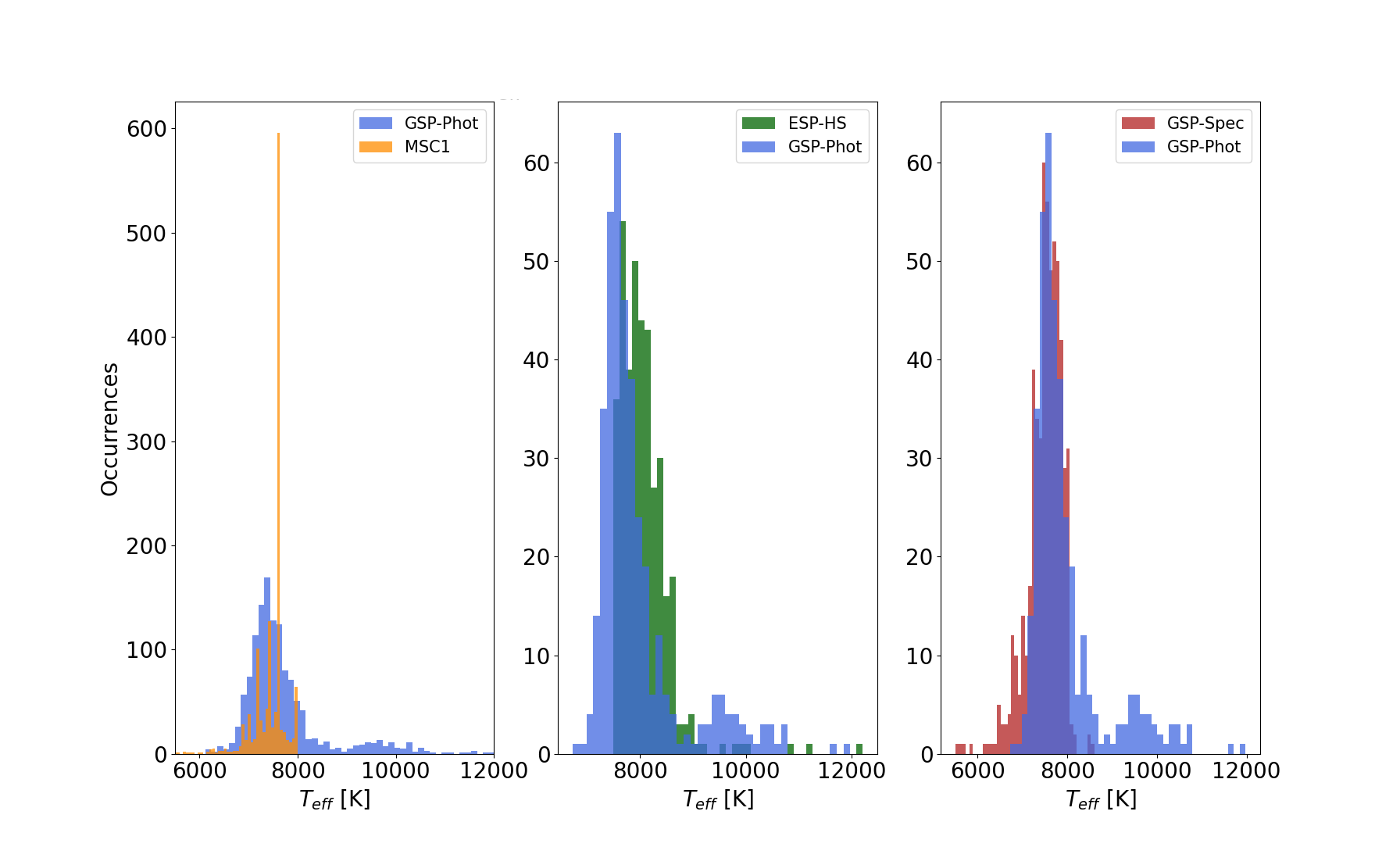}
   \caption{Comparison of the distribution of effective temperature in our sample of Am/Fm stars for different modules in \textit{Gaia} DR3. The GSP-Phot module is shown in blue in all three panels. Orange: assumption of binary star system. Green: assumption of hot star ($T_{\rm{eff}} > 7500~K$). Red: Same assumptions as GSP-Phot, but using medium resolution spectroscopy instead of photometry.}
   \label{Fig:GaiaTeff}%
\end{figure*}

\begin{figure}
    \centering
    \includegraphics[scale=0.2]{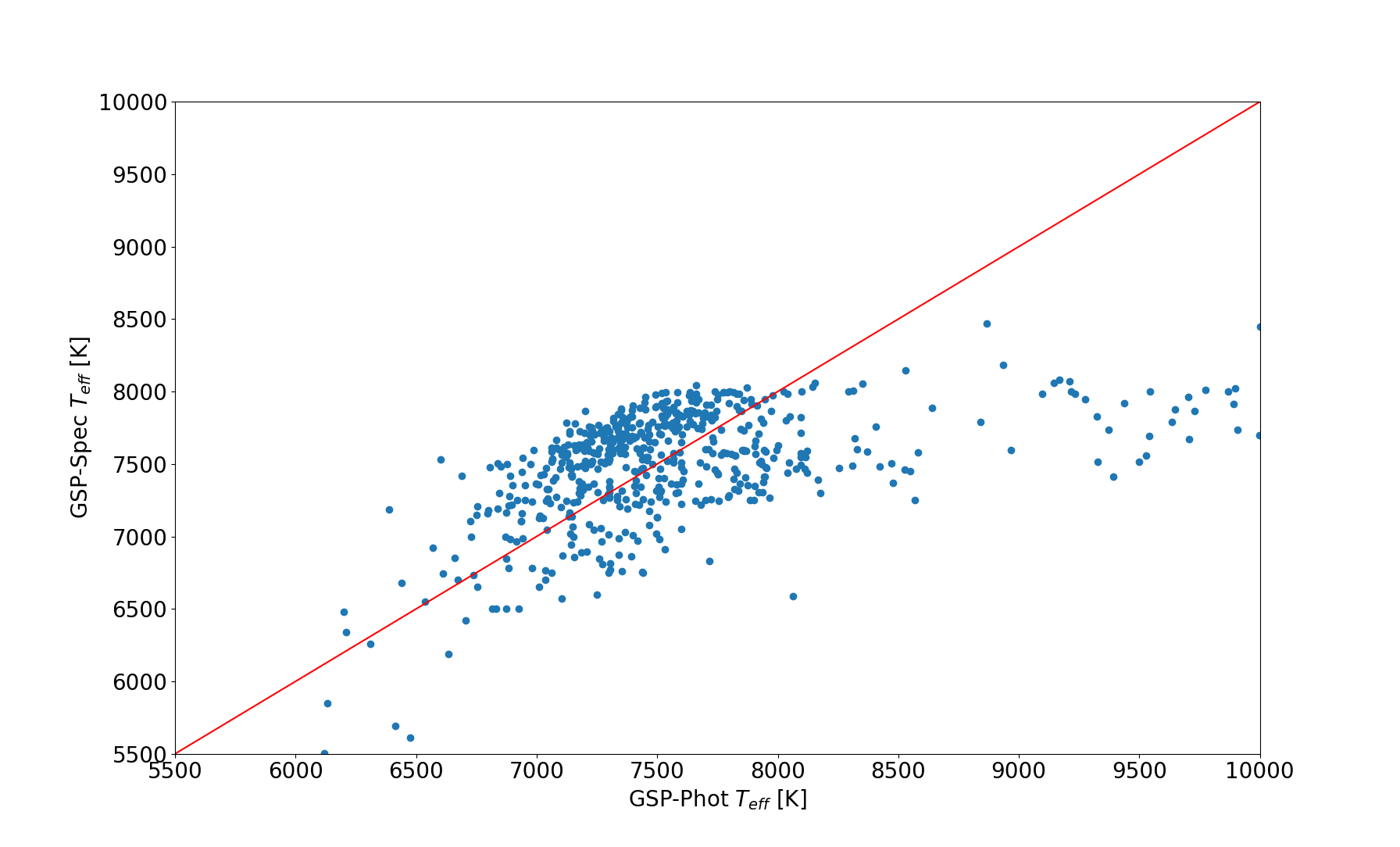}
    \caption{Comparison of the effective temperature from the GSP-Phot and GSP-Spec modules for stars present in both samples, highlighting the classification difference at higher temperatures.}
    \label{fig:TeffCompare}
\end{figure}

\begin{figure}
   \centering
   \includegraphics[scale=0.2]{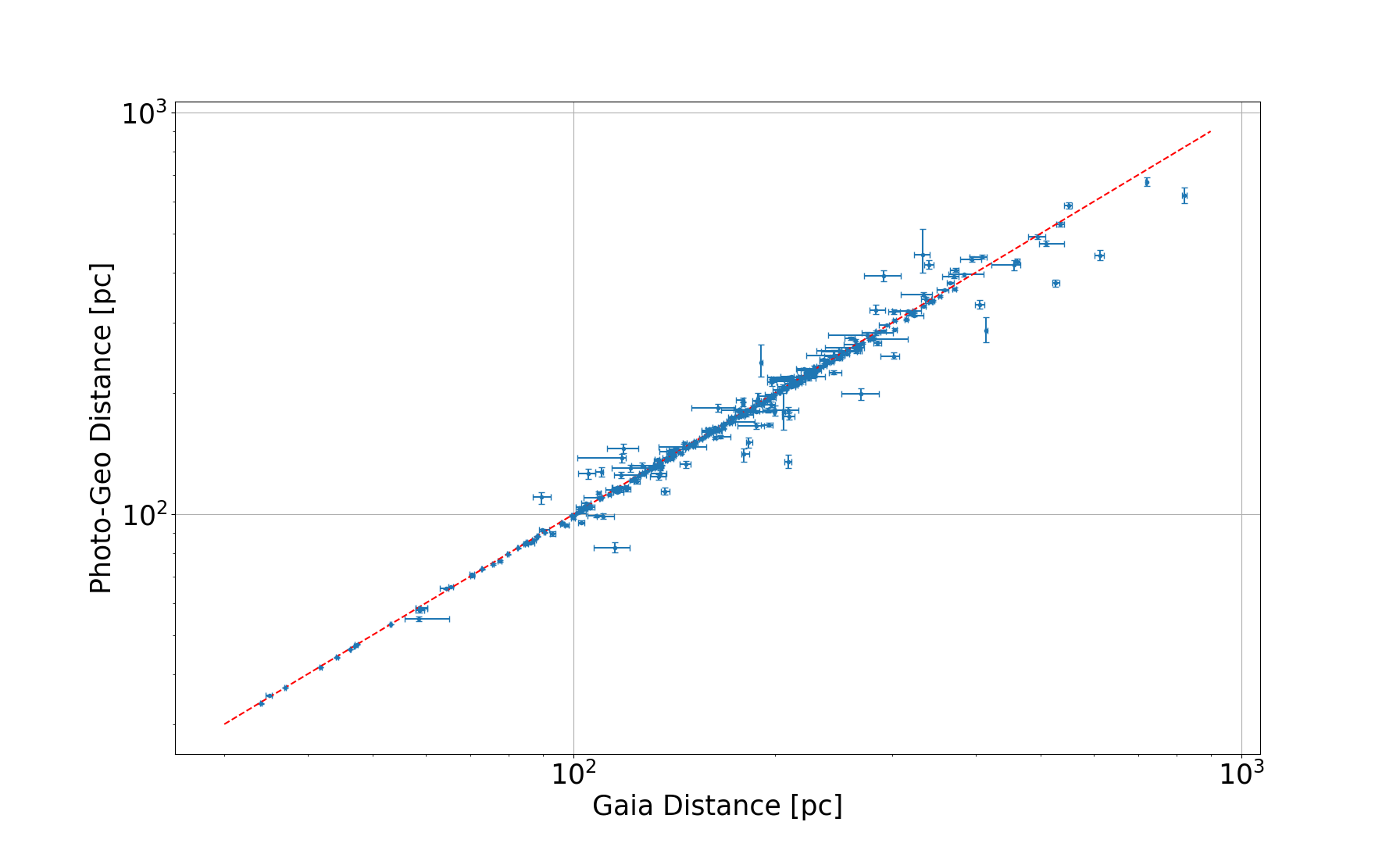}
   \caption{Comparison of the distance estimates in \textit{Gaia} DR3 from the GSP-Phot module and probabilistic inference from \cite{Bailer-Jones_2021}. The graph is shown in logarithmic scale, with the 1:1 relation highlighted by the red dotted line. Overall there is a good agreement between both methods, suggesting the GSP-Phot values and luminosity derived from these should be sound.}
   \label{Fig:BailerJones}%
\end{figure}

\subsection{\textit{Gaia} stellar parameters}\label{sec:Gaia}
In order to study our sample it is important to retrieve a homogeneous set of stellar parameters. To this effect, we consider data from \textit{Gaia}, which uses three instruments to observe the target stars \citep{GaiaInstruments}. The astrometric instrument measures the position and velocity of stars using parallaxes and proper motion. The satellite is also equipped with Blue and Red Photometers (BP/RP) and a Radial Velocity Spectrometer (RVS). The Astrophysical Parameters Inference System (Apsis), operated by the \textit{Gaia} Data Processing and Analysis Consortium, analyses data from these instruments and provides estimates of astrophysical parameters. The recent \textit{Gaia} DR3 is unprecedented in the number of data products made available to the science community and guarantees a homogeneous source of stellar parameter estimates. It therefore provides a unique opportunity to perform ensemble analysis of Am/Fm stars. The \textit{Gaia} Online Documentation provides a detailed overview of the different data products available and how these were derived \citep{GaiaDocumentation, ApsisI, ApsisII}, which we used to identify the stellar parameters of interest. The following modules were considered in our data selection process:
\begin{enumerate}
    \item \textit{GSP-Phot}: General Stellar Parameteriser from Photometry, using low-resolution spectra obtained from the BP/RP instruments to compute stellar parameters
    \item \textit{GSP-Spec}: General Stellar Parameteriser from Spectroscopy, similar to GSP-Phot but using medium-resolution spectra 
    \item \textit{MSC1}: the Multiple Star Classifier, which assumes the target is in a binary system.
    \item \textit{ESP-HS}: one of the Extended Stellar Parameterisers, which computes stellar parameters for hot stars, above 7500K.
    \item \textit{FLAME}: calculates luminosity, mass and age based on parallaxes and input from GSP-Phot and extinction map derived from DR2 photometry \citep{Lallement2019}.
\end{enumerate}

Since all modules overlap in their $T_{\rm{eff}}$ estimations, it allows us to assess whether the values are precise and can be used to perform ensemble studies. We investigated the effective temperature distribution of our stars as estimated by the first four modules, see Fig. \ref{Fig:GaiaTeff}, in order to determine the best possible choice. We eliminate the MSC1 since it leads to discrete temperatures dominating the distribution instead of a more continuous spread of values, the latter being expected as a more plausible representation of the sample. This unusual distribution is likely due to the method employed in the module \citep{GaiaOnlineDoc11}. The ESP-HS module is also eliminated due to an abrupt lower cutoff at 7500~K and does not fit the temperature distribution of known chemically peculiar Am/Fm stars. GSP-Phot is the only module providing an estimate for all stars in our sample and is a powerful tool for target selection and combined analyses \citep{Andrae2023}. It is also mostly consistent with GSP-Spec, as can be seen in Fig. \ref{fig:TeffCompare}. The major difference is the large number of stars with high $T_{\rm{eff}}$, which can be seen as a second hump in the distribution. As the only module showing this trend, it may suggest that the temperature of some stars may be overestimated in GSP-Phot because these higher temperatures correspond to B-type stars. Another explanation would be that these stars in reality are not Am/Fm stars, which would have to be investigated. We do not look further into this sample of stars and focus instead on those within the instability strip. Since the FLAME module computes the luminosity based on input from GSP-Phot in some cases, we decided for consistency to use these two modules to obtain stellar parameters of our sample. 

The uncertainties associated with the values have a tendency to be underestimated in the case of effective temperature and surface gravity. Comparisons made with external catalogues such as APOGEE, GALAH, RAVE, LAMOST, show that a median error of 110~K and 0.2~dex, respectively, for the former and latter parameter, should be used instead \citep{ApsisII}. On the other hand, luminosity could be slightly overestimated if the parallax is not well determined, or if the distance to the star as calculated by GSP-Phot is biased. As additional steps in the validation of the \textit{Gaia} data for our sample, we firstly computed the parallax fractional uncertainty for each star in our sample and confirm that these lie under 15\% except for a few outliers. This should in most cases ensure the luminosity is not overestimated \citep{ApsisI}. We then verified the distance estimates obtained from GSP-Phot in case these are used. Using \textit{Gaia} EDR3 parallaxes as well as colour information, \cite{Bailer-Jones_2021} used probabilistic inference \citep[see also][]{Bailer-Jones_2015} to determine new values for distances and confirmed that the values were generally of higher accuracy and precision through simulations and external validation. We downloaded these probabilistic estimates for our stars from the \textit{Gaia} Archive and compared them with distances from GSP-Phot, as shown in Fig. \ref{Fig:BailerJones}. Since we find that the values agree well, we assume that the data from the GSP-Phot and FLAME modules are reliable enough to use for analysis of stellar oscillations. 

\subsection{Renson catalogue}
Our target list is generated from the 4299 probable Am/Fm stars present in the Renson catalogue \citep{Renson}. This list also contains $\rho$ Puppis stars, included in our sample since they may be related to Am/Fm stars, which are labelled as $\delta$ Del. We used the coordinates of each star in the catalogue to perform a positional cross-match with the \textit{Gaia} Archive\footnote{\url{https://archives.esac.esa.int/gaia}} and MAST in order to obtain a consistent list of identifiers allowing for easy handling of subsequent data. When multiple matches were returned in MAST, we used the TESS Asteroseismic Science Operations Centre (TASOC)\footnote{\url{https://tasoc.dk}} database to identify the right target, as all but one turned out duplicate target IDs for the same star. In this way, it was possible to obtain the TESS Input Catalog (TIC) numbers and \textit{Gaia} DR3 identifiers for each object. At the same time, we verified the data availability for all stars and removed from our sample any target that did not have the required stellar parameters from \textit{Gaia} or photometric time series from TESS. This brought the final target list to 1276 Am/Fm stars to be analysed. 

Spectral classification of Am/Fm stars conventionally provides at least two spectral types, since the Ca II K-line appears to belong to an earlier spectral type than the metallic lines. In some cases, a third class can be assigned based on the hydrogen lines. Computing the numerical difference between spectral index obtained from the metallic lines ($m$) and the Ca II K-line ($k$) yields a measure of the stellar metallicism $\Delta$ \citep{Hou2015, Smalley2017}. Using this definition, larger values indicate stronger chemical peculiarities. Out of the 1276 stars in our analysis sample, we found almost 870 targets where two spectral types were provided, which we can use to study metallicism in Am/Fm stars and examine its possible effect on pulsation.

%--------------------------------------------------------------------
\section{Stellar oscillations}
In this section, we detail our methods for identifying and analysing stellar oscillations in our population of Am/Fm stars. Our goal is to understand the excitation mechanism in Am/Fm stars, specifically the role of He and its depletion, and derive key statistics regarding the pulsation characteristics of Am/Fm stars in this process.  

\begingroup
\setlength{\tabcolsep}{5pt}
\begin{table*}
\caption{Mean \textit{Q} constant estimates of the radial fundamental mode ($l=0, n=0$) and higher orders $n>0$ for A and F-type stars.}
\label{tab:Q}      
\centering          
\begin{tabular}{cccccccc}
\hline
\noalign{\smallskip}
$Q_0$ & $Q_1$ & $Q_2$ & $Q_3$ & $Q_4$ & $Q_5$ & $Q_6$ & $Q_7$ \\ 
\noalign{\smallskip}
\hline              
\noalign{\smallskip}
0.03290& 0.02521& 0.02021& 0.01682& 0.01440& 0.012 & 0.01113& 0.00996 \\
\noalign{\smallskip}
\hline                  
\end{tabular}
\tablefoot{Computed from the values provided in \cite{Fitch1981}.}
\end{table*}

\subsection{Variability determination}\label{sec:Contamination}
We computed amplitude spectra from the TESS light curves using the Lomb-Scargle periodogram \citep{Lomb, Scarlge}, a common algorithm for the analysis of unevenly spaced data. The spectra were analysed to determine the variability of our stars. This was done over two iterations with successively stronger filtering constraints and visual inspection of spectra and light curves between each phase. The latter focused on visually differentiating significant pulsation peaks from random occurrences. In the first step, we determined a mean noise level for each star by considering the frequency range 250--360~d$^{-1}$, where little astrophysical signal is expected (except for roAp stars, which were not found in the sample). This was used to verify if the frequency of highest amplitude was above the commonly used significance threshold of $S/N \geq 4$ \citep{Breger} for identifying significant pulsation frequencies. Visual inspection of some of the spectra from stars labelled as variable showed that this detection threshold was not successful at isolating the truly variable stars. Most contained high amplitude peaks near 0.7~d$^{-1}$, introduced by data gaps, or spurious frequencies, due to random noise. A second filtering stage was therefore required, where we adopted the stricter condition of $S/N \geq 10$ advocated by \cite{Murphy2019} when looking at the strongest peak. Great results were achieved with this method, successfully separating variable Am/Fm stars from non variable ones. In addition, to ensure accurate classification, we visually inspected all stars. It is important to note that some stars classified as non-variable may exhibit variability below our detection threshold ($S/N \geq 10$). 
The $S/N = 4$ was also computed by using a sliding window to determine local values of the noise across the spectrum. The window size is set to respectively 2, 10 and 50 $\mathrm{d}^{-1}$ for the intervals 0--10, 10--100 and 100--360 $\mathrm{d}^{-1}$ to be more sensitive at low frequencies. Although this is not used in the analysis, it is included in visualisations of amplitude spectra to help the eye with identifying areas with higher noise levels.

Out of the 1276 stars analysed, 637 variable candidates remained. These were then manually inspected in order to classify their variability type. We defined the following three possible variability classes:
\begin{itemize}
    \item Pulsating stars
    \item Eclipsing binary systems 
    \item Modulated light curves and other variation
\end{itemize}
Our aim was then to inspect the amplitude spectra and light curves in detail to identify features that would place these stars in one (or more, as binaries can also exhibit pulsations) classes. We created a visual Am/Fm catalog, summarising key stellar parameters and metadata from \textit{Gaia} and TESS as well as displaying the time series and amplitude spectra for each star. Targets for which g-mode or p-mode pulsations were identified, are labelled as pulsators. Since low-frequency g-modes and p-modes originate from different mechanisms and regions of the star, they relate to different properties and can readily be distinguished from one another. We adopted a soft boundary of 4~d$^{-1}$ to separate these, allowing for frequencies to extend from the g-mode domain into low frequency p-modes and vice versa, since low order g-modes can oscillate at more than 4~d$^{-1}$. To help assess whether frequencies in this grey zone are g-modes or p-modes, we considered the predicted location of the radial fundamental mode for $\delta$ Scuti stars (see Section \ref{sec:Radial}), below which only mixed and g-modes are present. A few stars exhibit only one high-amplitude peak in the p-modes regime at frequencies too high to be attributed to rotational modulation or binarity. We categorised these as p-modes as well but recommend more detailed analyses for further confirmation. Some pulsating stars might also exhibit ellipsoidal variation and/or rotational modulation. It is important to note that we cannot exclude the possibility of misclassifying g-modes as rotational modulation or vice versa. Should our catalogue be used to aid studies about the incidence of hybrid stars we recommend more thorough analyses of the low frequency peaks in the case of the hybrids but also of the p-mode pulsators. Some of the low-frequency peaks might not originate from high-order g-modes, while some p-mode pulsators could also exhibit genuine high-order g-modes.

%We classify a star as showing g-modes if it has at least two peaks that are not related through harmonics.
We searched for stars exhibiting binarity through eclipses or ellipsoidal variation, the latter creating a distinct \textit{M}-shape in the light curves \citep{Greene2001}. A literature search using TASOC was conducted for these candidates to accurately classify known binary systems in our sample, even when our data showed no direct evidence. Stars with clear variability not fitting these two and above categories were labeled as modulated or `other'. These stars typically show trends in their light curves linked to modulation by stellar spots. Due to the potential overlap between ellipsoidal variation and rotational modulation, often leading to ambiguous classification, we combined non-pulsating modulated stars and ellipsoidal variables under the category `Other variability', while eclipsing binaries were assigned their own distinct group. The `Other variability' group also comprises a few eccentric binaries showing a characteristic `heart-beat' shape in their light curves \citep[e.g.,][]{Thompson}. We cannot rule out the possibility that some g-mode pulsators with unresolved frequencies may have been misclassified as rotational modulation, and thus identified as `Other variability'. This issue can only be solved with additional data.

This classification process was carried out for all  stars over multiple iterations, ultimately leading to a well-defined sample of pulsating Am/Fm stars. Because TESS has a wide field of view and large pixels, there is a high risk of contamination of a target's light curve from another closely neighbouring star. We retrieved a contamination estimate from TASOC for each star and looked up a sky image to see if they had large values. We downloaded light curves for the 2--4 nearest stars of similar magnitude to the target Am/Fm star and compared the amplitude spectra. Several cases showed identical frequency profiles, with the star having larger amplitudes thus being ruled as the variable star. In cases where this was the neighbouring stellar object, we eliminated the Am/Fm target from our sample. In total we found 18 contaminated cases, which provide us with a first estimate of the classification uncertainty and brings the total number of variable sources in our sample down to 619 $\pm 18$ stars.

\subsection{Frequency extraction}
Pre-whitening was carried out with the Extraction of CoHerent Oscillations (ECHO) program, developed at the Stellar Astrohysics Centre at Aarhus University. ECHO computes an amplitude spectrum using a fast Lomb-Scargle algorithm \citep{PressRybicki} and performs iterative extraction of frequency, amplitude, and phase according to the algorithm presented by \cite{Frandsen}, with associated uncertainties derived as described in \cite{Montgomery}. Additional information about the ECHO pipeline can be found in \cite{Antoci2019}. We allowed for the maximum amount of frequencies to be extracted, but selected a strict false-positive rate of $p=0.0027$ and stopped the pre-whitening after 7 consecutive extraction failures. Because iterative pre-whitening may introduce artificial signals that become statistically significant \citep{Balona2014}, a criterion for the ratio of the extracted frequency's amplitude and the one present in the original spectrum is implemented:
\begin{equation}
    \alpha < \frac{A_{extracted}}{A_{original}} < \frac{1}{\alpha}
\end{equation}
Based on simulations of \textit{Kepler} data it is recommended to be set as $\alpha=0.75$ \citep{Antoci2019}. We found that in some rare cases the $\alpha$ threshold had to be relaxed in order to extract frequencies correctly for stars where pulsations were clearly identifiable by eye, as seen in the example in Fig. \ref{Fig:ExampleNoExtract}. 

\begin{figure*}
    \centering
    \includegraphics[width=\textwidth]{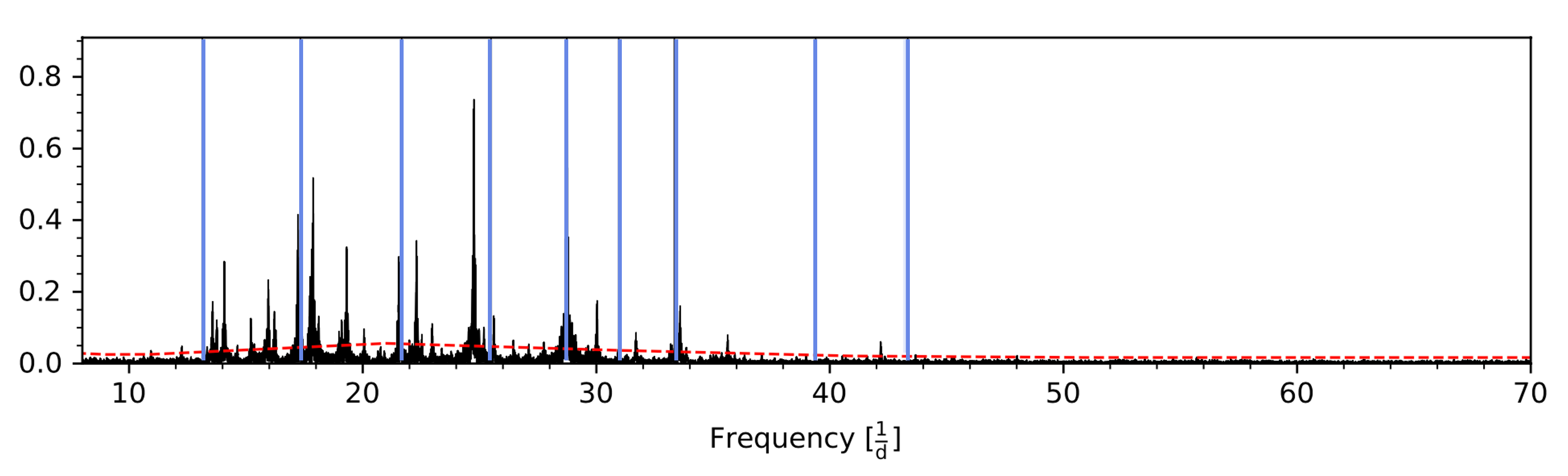}
    \caption{Example amplitude spectrum of a star for which frequency extraction did not yield any results. The vertical blue lines spanning the entire window show the location of the predicted radial orders. The red dashed line shows the $S/N = 4$ line, computed as a sliding window. This star was upon visual inspection included in the sample and conditions were loosened in order to extract frequencies for it.}
    \label{Fig:ExampleNoExtract}
\end{figure*}

A crucial step following pre-whitening is identifying the range of excited modes, which requires the identification of linear combination frequencies in the extracted signals. These linear combinations are not intrinsically excited stellar oscillation modes and would lead to overestimating the range of excited pulsations. With the development of space-borne missions observing pulsating stars, which allow for detection of much weaker oscillation frequencies compared to ground telescopes, combination frequencies have become increasingly important to identify and remove from further analysis. As a first measure, we perform a search for combination frequencies of order $O=2$ ($2f_i$, $f_i \pm f_j$), involving at most two parent frequencies, as higher order combinations are unlikely to occur if the lowest order is not detected \citep{Papics}. A second iteration dedicated to higher orders is carried out for stars where there are positive matches. The outcome of the ECHO pipeline is a list of oscillation frequencies with associated amplitudes and phases, as well as their uncertainties, for each star. 

\subsection{Radial mode excitation}\label{sec:Radial}
Due to the short time series of a significant number of stars, impeding the mode identification of g-modes in our list of $\gamma$ Doradus stars, we limit our further detailed analysis to p-modes in $\delta$ Scuti or hybrid stars. We focus our work on the study of radial modes, with degree $l=0$, attempting to identify the radial fundamental mode and any higher orders. The pulsation constant \textit{Q} \citep{Petersen} can be used to predict the period of radial orders for a given star \citep{Breger} according to the following equation: 
\begin{equation}
    \log Q = -6.454 + \log P + 0.5\log g + 0.1 M_{bol} + \log T_{\rm{eff}}
    \label{eq:Q}
\end{equation}
where \textit{P} is the pulsation period (1/frequency), $\log g$ is the surface gravity and $T_{\rm{eff}}$ is the effective temperature. Recent studies have made use of this relation to observational analysis of stellar pulsations \citep[e.g.][]{Zwintz2020} and modelling of convection and overshoot in pulsating stars \citep{Lovekin}. The bolometric magnitude $M_{bol}$ can be computed from it's absolute magnitude $M$, estimated from the FLAME module in \textit{Gaia} DR3, by adding the following bolometric correction \citep{Reed1998}:
\begin{multline}
    M_{bol} = M -8.499\cdot\log(T_{\rm{eff}})^4 + 13.421\cdot\log(T_{\rm{eff}})^3 \\ 
    - 8.131\cdot\log(T_{\rm{eff}})^2 - 3.901\cdot\log(T_{\rm{eff}}) - 0.438
\end{multline}
This produces an expression capable to predict the period of radial orders depending only on the effective temperature, surface gravity and $Q$ constant. Values for the latter have been derived for low degree p-modes, from the radial fundamental up to the seventh overtone, for different stellar masses \citep{Fitch1981}. From these values, we compute a mean $Q$ constant, covering the entire A \& F mass range, for each order and present them in Table \ref{tab:Q}. In recent years, new simulations of the pulsation constant have been made \citep{Lovekin}. Good agreement was found for $T_{\rm{eff}} \geq 6760$~K, covering the temperature range of interest for our sample, whilst new values are proposed below this threshold. We use the previously obtained values for the purpose of this study.

From equation \ref{eq:Q}, we compute the expected period of radial modes of different orders, and get their frequencies, for each star. Uncertainties for these frequencies are derived from Monte Carlo simulations. We sample the stellar parameters from a Gaussian distribution with mean value corresponding to the estimates in \textit{Gaia} DR3 and standard deviation the associated uncertainty (see Section \ref{sec:Gaia}). These are inserted into equation \ref{eq:Q}, providing a value for the frequency, and repeat this process over 10,000 iterations. We compute the median of all outcomes as our final value, and use the standard deviation as our uncertainty. We then inspected the amplitude spectra of all pulsating stars, highlighting the predicted frequencies obtained from $Q$, as well as the extracted and combination frequencies. A radial order (\textit{n}) is considered excited if there are any frequencies at higher values. Orders below the first radial order to show significant frequencies are not considered excited. This provides an estimation of the number of excited orders in each star, however no specific mode identification is carried out. In some cases, we observe gaps between excited radial orders, which could be intrinsic to a single star or indicate a binary system with two pulsating components. Interestingly, excitation models, such as those presented by, e.g. \citet[][]{Antoci2019}, also show similar gaps between low and intermediate excited radial orders.

%--------------------------------------------------------------------
\section{Results}

\begingroup
\setlength{\tabcolsep}{4pt}
\begin{table}
\caption{Summary statistics of the variability within our sample of Am/Fm stars.}
\label{tab:AmSummary}      
\centering          
\begin{tabular}{ccc}
        \hline
        \noalign{\smallskip}
        \multicolumn{3}{c}{TESS \& \textit{Gaia}: 1276} \\
        Type & Occurrence & Fraction \\
        \noalign{\smallskip}
        \hline
        \noalign{\smallskip}
        Constant & 649 & 51\% \\
        Pulsating Am/Fm \& $\rho$ Puppis & 318 & 25\% \\
        Pulsating Am/Fm  & 261 & 20\% \\
        Eclipsing binaries & 93 & 7\% \\ 
        Other variability & 210 & 17\% \\
        \noalign{\smallskip}
        \hline            
        \noalign{\smallskip}
        \multicolumn{3}{c}{Pulsation incidence} \\
        \noalign{\smallskip}
        \hline            
        \noalign{\smallskip}
        $\gamma$ Doradus & 32 & 10\% \\
        $\delta$ Scuti & 172 & 54\% \\
        Hybrids & 114 & 36\% \\
        \noalign{\smallskip}
        \hline            
\end{tabular}
\tablefoot{The pulsation statistics of our oscillating targets are also shown. We find a 20\% pulsation incidence in Am/Fm stars.}
\end{table}
\begin{figure*}
    \centering
    \includegraphics[scale=0.4]{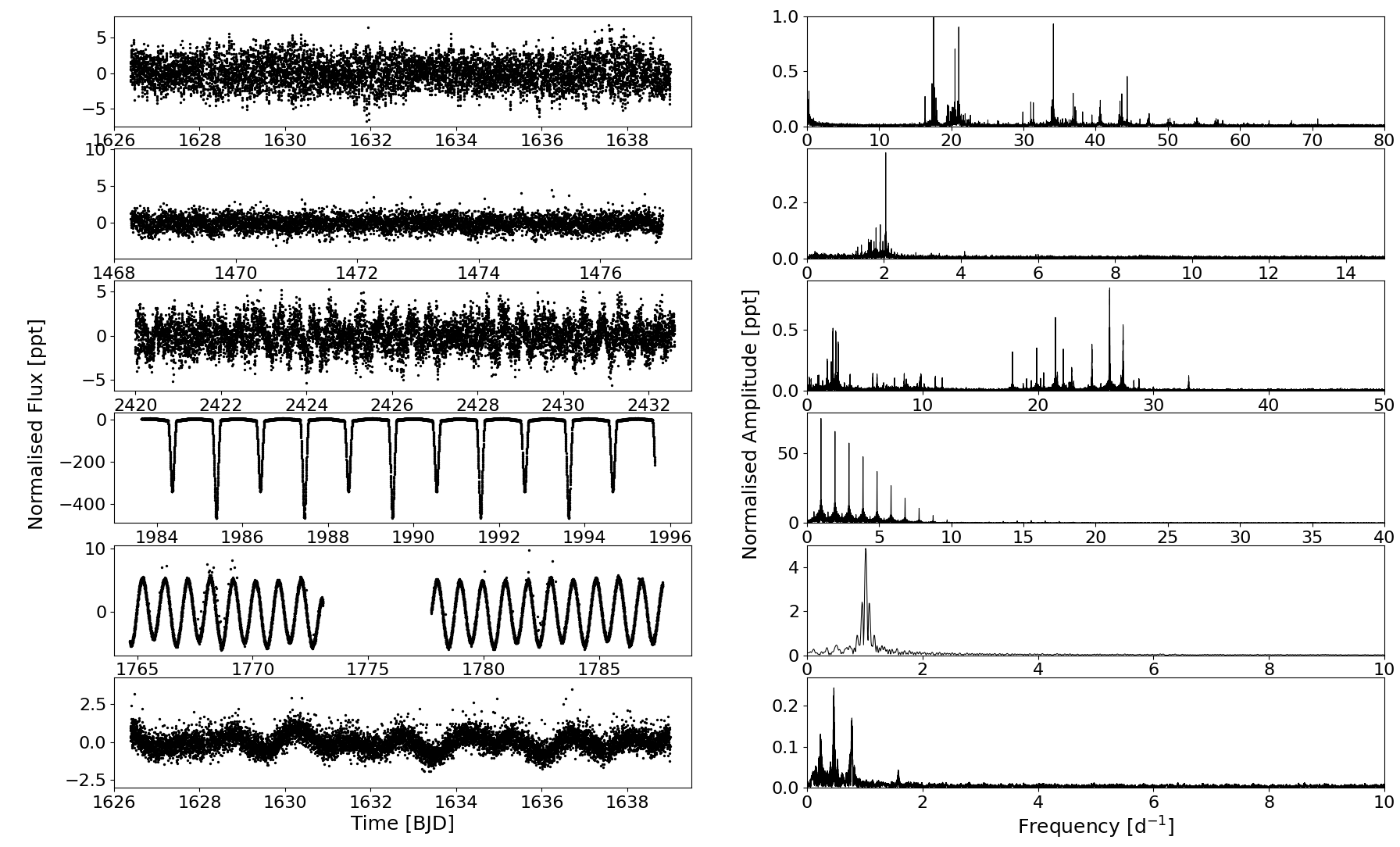}
    \caption{Light curves and amplitude spectra of Am/Fm stars belonging to different variability classes. The time series only span half a sector for visual clarity. The frequency interval is also reduced to highlight the features present in the spectra, although these extend to 360~d$^{-1}$. From top to bottom: TIC 384069471 ($\delta$ Scuti); TIC 279090002 ($\gamma$ Doradus); TIC 273754456 (Hybrid); TIC 101675455  (Eclipsing binary); TIC 191460663 (Ellipsoidal variability; other variability class); TIC 401667994 (Rotational modulation; other variability class).}
    \label{Fig:Variability}%
\end{figure*}
We present in this section the results of our analysis of pulsation types in Am/Fm stars. We compiled one-page summaries for each star, with information about stellar parameters, the number of TESS sectors available, and most importantly the light curve and amplitude spectrum of the star. The latter was also divided into smaller intervals to allow more detailed inspection of some regions of interest. From these we constructed Am/Fm catalogs, grouping stars into the following categories:
\begin{itemize}
    \item $\gamma$ Doradus pulsators
    \item $\delta$ Scuti pulsators
    \item Hybrid pulsators
    \item Eclipsing binaries
    \item Constant
    \item Other variability
\end{itemize}
These catalogs contain a wealth of information while providing readers with a good overview due to its simple layout. They are the starting point of our analysis of variability and radial order excitation among the Am/Fm population, facilitating greatly the analysis of a large number of Am/Fm stars, and have been made available online \footnote{\url{https://github.com/Durfeldt/AmStars_Catalogues}} in the hopes that they may aid the community in picking out interesting targets for further study. In order to give readers an idea of the variability types, we show some example light curves and amplitude spectra of each class in Figure~\ref{Fig:Variability}. Moreover, we provide a summary table of all our Am/Fm stars, containing useful identifiers for cross-matching as well as stellar parameters, variability type and a flag indicating whether the star is a known binary or not. If applicable, we give an estimate of the frequency range of g-mode and p-mode pulsations and include our estimate for the number of excited radial orders for p-modes, which will be described in greater detail in Section \ref{sec:RadialResults}. Table \ref{tab:AmStars} is an excerpt from the full data frame, available online at the CDS.

\begin{figure*}
    \centering
    \sidecaption
    \includegraphics[width=12cm]{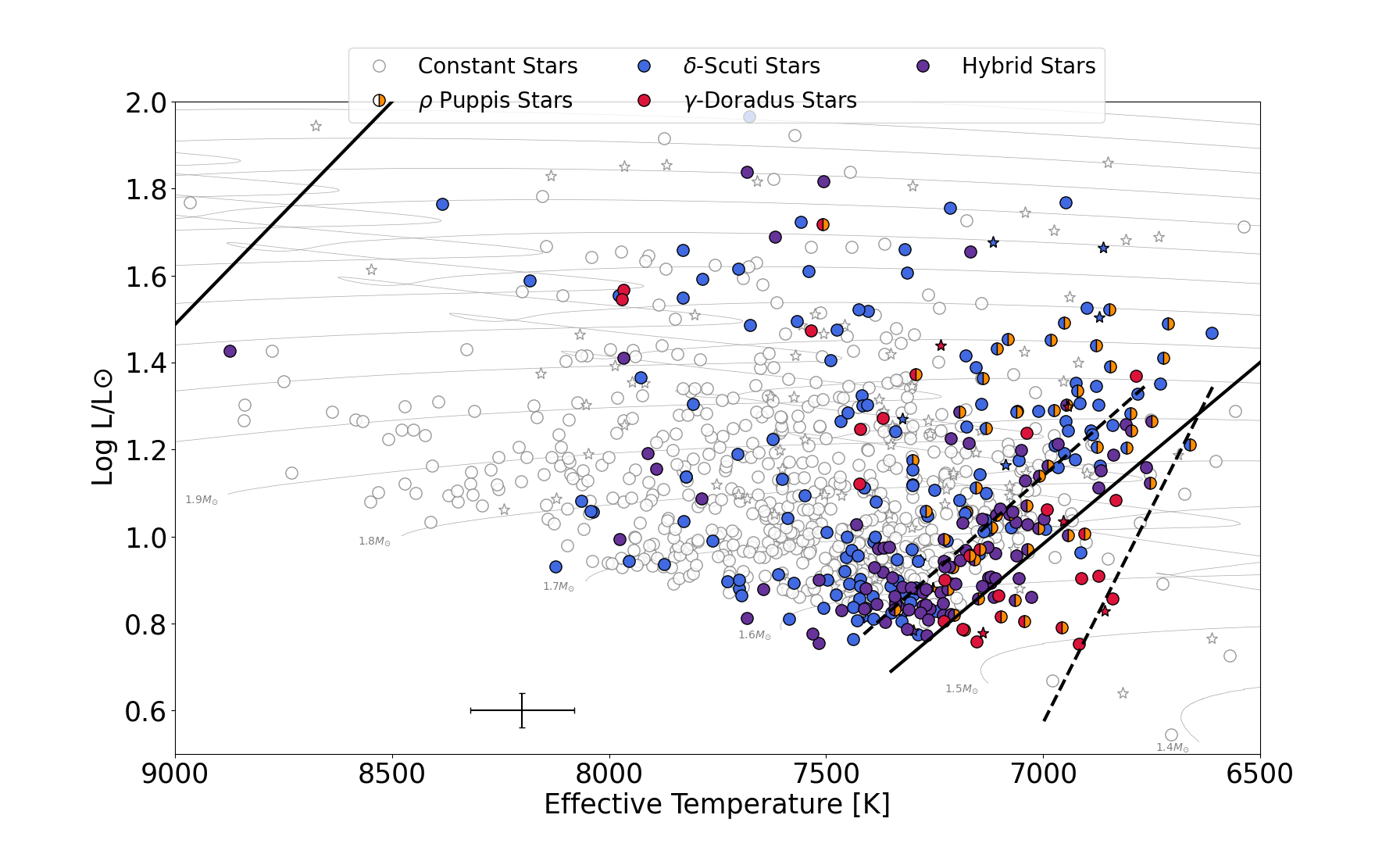}
    \caption{HR diagram containing all 1276 Am/Fm stars studied using TESS and \textit{Gaia} data. The typical error bars for our stars are shown in the lower left part of the plot. Constant stars are shown in white. Binary systems are highlighted with a star-shaped symbol. The pulsation type is colour-coded according to the following colour scheme; red: $\gamma$ Doradus, blue: $\delta$ Scuti, purple: hybrids. The $\rho$ Puppis stars are also indicated with an orange half-circle. We show the $\gamma$ Doradus instability strip from \cite{Dupret2005} in black dashed lines for $\alpha=2$. The $\delta$ Scuti instability strip is from \cite{Murphy2019}, and is plotted with solid black lines. We include evolutionary tracks to facilitate the interpretation of the figure, which are based on Warsaw-New Jersey models assuming no rotation and solar metallicity \citep{Dziembowski1977, Pamyatnykh1998, Pamyatnykh1999}.}
            \label{Fig:HRD}%
    \end{figure*}
    
\subsection{Am/Fm population statistics}
Starting off with the 4299 Am/Fm stars in the Renson catalogue, we were able to retrieve 2 minute cadence TESS light curves and full \textit{Gaia} stellar parameter estimation for 1276 stars, i.e. 30\% of the initial sample. Many targets were excluded due to the absence of values in the FLAME module. This still provides a large enough sample to determine some characteristics of the Am/Fm population. Table \ref{tab:AmSummary} summarises the outcome of our variability classification. The majority of Am/Fm stars are non-pulsating, with 1\% of our sample showing no features in the amplitude spectrum. We find $261 \pm 18$ Am/Fm stars with clear stellar oscillations, where the uncertainty is derived from the number of contaminated stars found in the sample (see Section \ref{sec:Contamination}). In total, 75 pulsating stars were flagged with a contamination value greater than 0.1 in TASOC, out of which  turned out to truly be contaminated. These are omitted from the statistics in Table \ref{tab:AmSummary}. Additionally, 12 stars were identified as both pulsating and eclipsing binaries and are therefore included in the numbers for both classes.

This result leads to a pulsation fraction of 20\% in our Am/Fm population. Within our full list of pulsating stars, we find 90\% to be p-mode pulsators. With 4 hybrids and  pure $\gamma$ Doradus stars, slightly less than half of our sample also hosts g-mode pulsations. Variations unlikely to be caused by stellar oscillations were observed in 210 stars, categorised as `other variability'. These variations align with patterns of rotational modulation or ellipsoidal variability. When feasible, we validated binaries through a literature search. Nevertheless, this type of variability is outside the scope of this paper and was not examined in further detail. Furthermore, we found 93 eclipsing binary systems, leading to an eclipsing binary incidence of \% in our sample. We emphasise that to fully understand the incidence of binarity in Am/Fm stars, in-depth analyses are necessary. Our classification should be viewed as a starting point only.
    
\textit{Gaia} DR3 provides for the first time access to tables of non-single stars (NSS) \citep{GaiaDR3_NSS}, which could aid us in identifying more possible binary systems in order to better constrain the binary fraction in Am/Fm stars. However, only a very small fraction of objects in our sample are found in these tables, with a few RV estimates being available. The analysis in the NSS tables is carried out only for a small portion of the available data for DR3, focusing on unresolved astrometric binaries and spectroscopic binaries with at least 10 transits and effective temperatures between 3875~K and 8125~K \citep{GaiaNSS}. Stars that do not meet these requirements are excluded despite being suspected binaries. Due to the limited additional information to be retrieved, we consider that the NSS tables in their first version do not contribute significantly to the study of Am/Fm binaries, although this data may prove useful in the future.

\begin{figure*}
    \centering
    \sidecaption
    \includegraphics[width=11cm]{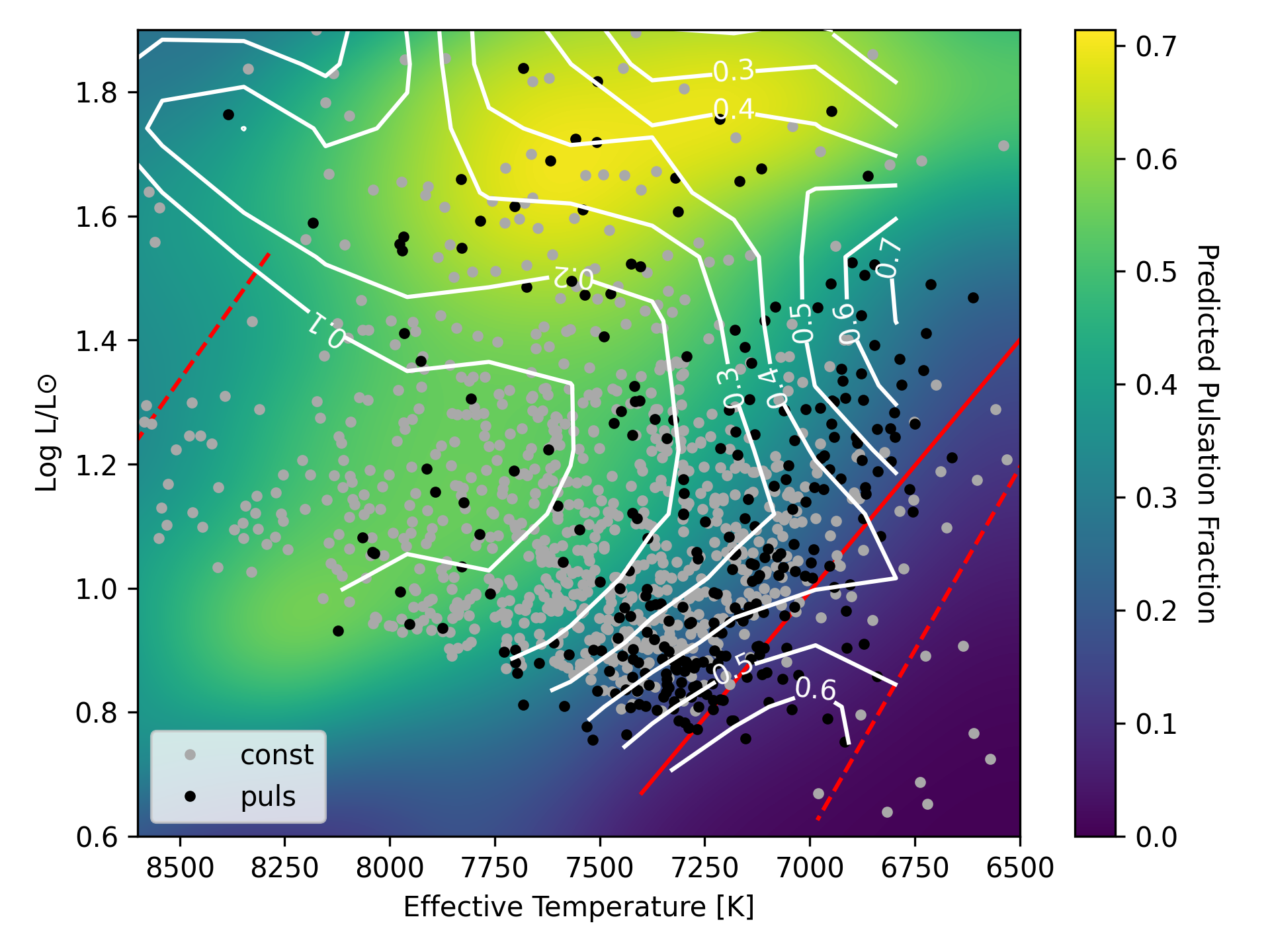}
    \caption{The heat map is the predicted pulsation fraction for all A and F stars from \citet{Murphy2019}. The black dots are Am/Fm stars found to exhibit $\delta$\,Sct pulsations, while the those without known pulsations are shown in light grey. The white contours indicate the observed Am/Fm star pulsator fraction. Theoretical instability strip boundaries from \citet{Dupret2005}, for $\alpha=2$, are show as dashed lines, while the red-edge given by \citet{Murphy2019} is shown as a solid red line. The typical uncertainties for $T_{\rm{eff}}$ and luminosity are the same as in the previous figures. }
    \label{fig:heatmap}
\end{figure*}

\begin{figure*}
    \centering
    \sidecaption
    \includegraphics[width=12cm]{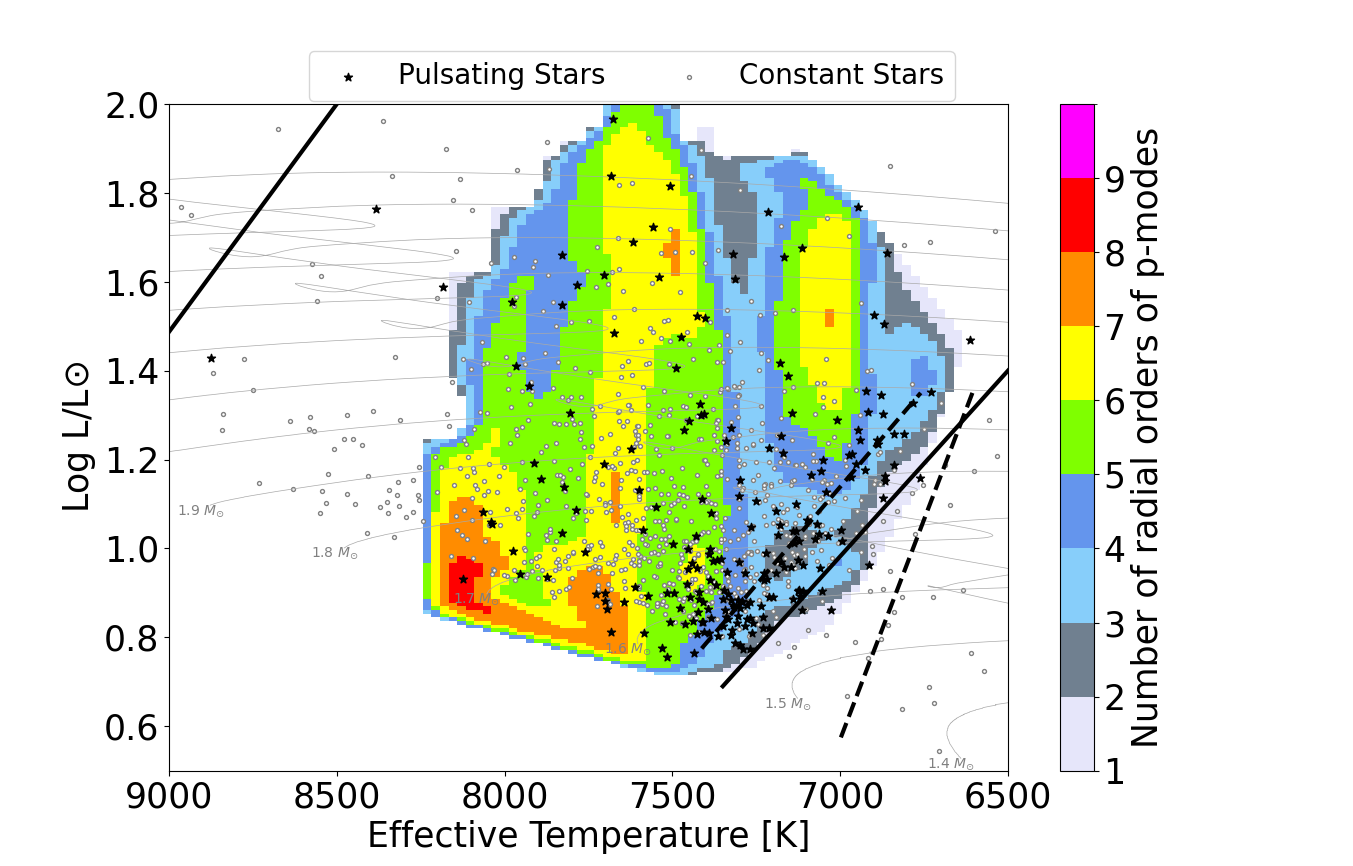}
    \caption{Distribution of radial order excitation in the HR diagram for our pulsating Am/Fm stars. The map shows the estimated number of radial orders excited. Non pulsating stars are shown in white circles, while our sample of pulsating stars are seen as black stars. The instability strips and evolutionary tracks are the same as those shown in Fig. \ref{Fig:HRD}.}
    \label{Fig:OrdersHeatmap}%
\end{figure*}

\begin{figure*}
    \centering
    \sidecaption
    \includegraphics[width=12cm]{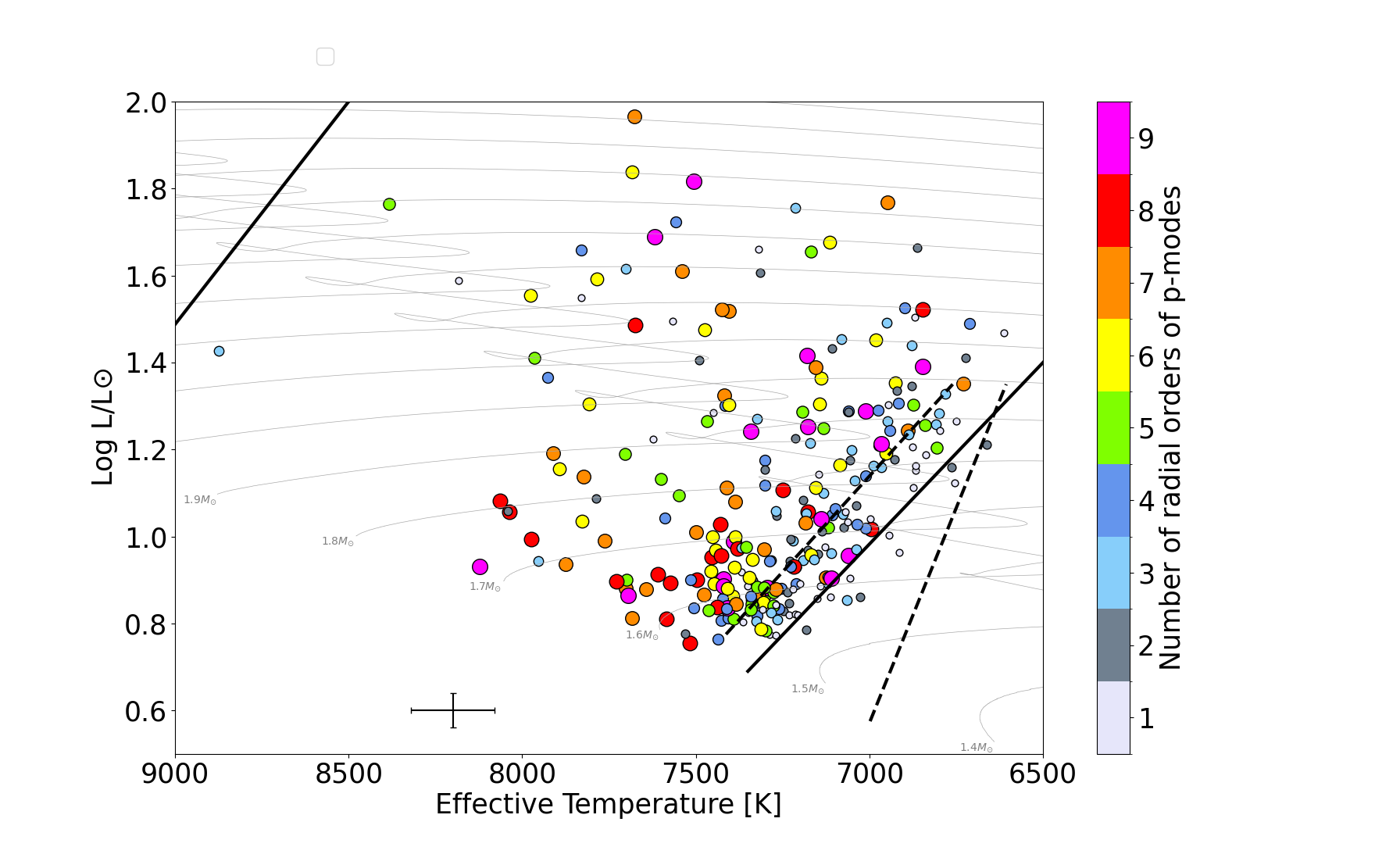}
    \caption{HR Diagram showing the observed number of excited radial orders for each star hosting p-modes. The instability strips and evolutionary tracks are the same as those shown in Fig. \ref{Fig:HRD}. The typical uncertainties for $T_{\rm{eff}}$ and luminosity are shown in the lower left part of the plot.}
    \label{Fig:OrdersScatter}%
\end{figure*}

\subsection{Pulsation incidence}\label{sec:RadialResults}
In Fig. \ref{Fig:HRD} we show our sample of stars in the HR diagram using the \textit{Gaia}-derived stellar parameters and their recommended uncertainties (see Section \ref{sec:Gaia}). All are colour-coded according to their pulsation type. Binary systems, including the eclipsing binaries found from the variability analysis and other systems known from literature, are highlighted by a star-shaped symbol. The $\rho$ Puppis stars are given an orange label to distinguish them from the rest of the sample. Excluded from this figure are a handful of outliers with effective temperatures above 9000~K, some of which host $\delta$ Scuti pulsations. We choose to exclude these stars until external confirmation of the effective temperatures, as they belong to the second hump in the temperature distribution for the GSP-Phot module (see Fig. \ref{Fig:GaiaTeff}, not present in GSP-Spec, which uses medium-resolution spectroscopic data) and $T_{\rm{eff}}$ therefore seems to be overestimated in these cases. One explanation could be that these stars are in binary systems and that the companion contributes to the average temperature significantly. The presence in a binary system could also imply that the pulsations come from the companion star. Unfortunately, no further information is available on these targets.

Most of the pulsating Am/Fm stars are located within the expected regions of instability, with hybrid stars mostly found in the overlap region between the $\delta$ Scuti and the $\gamma$ Doradus stars. Interestingly the $\delta$ Scuti Am/Fm stars tend to occupy the area near the red edge of the instability strip. This trend correlates well with previous results showing a narrower instability strip for He depleted pulsators and the blue edge being predicted to lie at cooler temperatures \citep{Murphy2020}. Moreover, there is a clear concentration of pulsating Am/Fm stars near the zero-age main sequence (ZAMS), suggesting they might show this behaviour predominantly at a young age. On the other hand, most $\rho$-Puppis pulsators are located near the terminal-age main sequence compared to pulsating Am/Fm stars, supporting the idea that they evolve from Am/Fm stars.

A direct evaluation of the effect of the Am/Fm peculiarity on pulsations is now possible. This is because the pulsation properties of $\delta$\,Sct stars are better understood as a function of temperature and luminosity, thanks to \textit{Kepler} photometry and \textit{Gaia} parallaxes \citep{Murphy2019}, and because TESS gives us a homogeneous view of the pulsations of Am/Fm stars. Where ground-based observations had previously been limited by Earth's atmosphere and the day-night cycle, observations of over 1000 Am/Fm stars with TESS constitute a unique dataset with which to make a meaningful comparison. 
While it is clear that some $\delta$\,Sct stars exist beyond the theoretical blue-edge of the instability strip \citep{Bowman2018}, the pulsator fraction in that area is low: there are also many non-pulsators. Similarly, while many $\delta$\,Sct stars exist at or beyond the red edge, these are overwhelmed in number by non-pulsators \citep{Murphy2019}. Following the method of \citet{MurphyLambdaBoo} developed to find the pulsator fraction of chemically peculiar $\lambda$\,Boo stars, we use the map of the \textit{Kepler} pulsator fraction in temperature and luminosity to create a model of the likelihood that a random star in our sample will pulsate, given its temperature, luminosity, and the uncertainties. Our grid cells are $\sim$200\,K in temperature and $\sim$0.1\,dex in luminosity, similar in size to our observational uncertainties, and the predictive model uses the pulsator fraction in neighbouring cells to infer model uncertainties. We correct for the sensitivity difference of the \textit{Kepler} and TESS spacecraft in exactly the same way as \citet{MurphyLambdaBoo}. 
It is clear from Fig.~\ref{fig:heatmap} that the distribution of pulsating Am/Fm stars is different from that given by the general population of pulsating A-type stars. The fraction of pulsating Am/Fm stars is very high close to the  \citet{Murphy2019} red edge of the observed instability strip, where the general A-star fraction is quite low (see Fig. 12 in \citet{Murphy2019}). Pulsational analyses using models with time-dependent non-local convection treatment similar to \citet{Antoci2014, Antoci2019}, which also allow for helium depletion in the outer envelope, are required to understand this trend.

\subsection{Radial order excitation}
We carried out an analysis of radial mode excitation for our 286 p-mode pulsating objects found in our sample, including Am/Fm stars as well as a few $\rho$ Puppis stars, according to the method described in Section \ref{sec:Radial}. A star is assigned a label, ranging from 0 (no p-modes but low-radial order g-modes and/or mixed modes) to 8, depending on how many radial orders are found to be excited. Some targets show more than eight radial orders excited, which we denote with 9. Only few stars have pulsation frequencies below the radial fundamental mode, suggesting that the frequencies we originally associated with p-modes instead must be low-order g-modes or mixed modes. The number of excited radial orders is given for each star in Table \ref{tab:AmStars}. In Fig. \ref{Fig:OrdersHeatmap} we show a radial order excitation map using the information from individual stars, which is also displayed in Fig. \ref{Fig:OrdersScatter} for clarity. The colour-coding represents the number of radial orders excited. To handle the unevenly distributed stars, we employ the Delaunay triangulation to interpolate data points over a mesh, followed by Gaussian smoothing to refine the results. The window settings define a temperature range of 400~K and a luminosity range of 0.4 in log scale, with a grid resolution of 200 points for both $T_{\rm{eff}}$ and logarithmic luminosity. At least three data points are needed for the triangulation to be valid. Gaussian smoothing is applied with a sigma value of 2 to produce a continuous and smoothed data representation.

In order to assert the significance of the excitation map, we also show the same region of the HR diagram in Fig. \ref{Fig:OrdersScatter}  with each pulsating star represented instead. We note a larger number of excited radial orders near the ZAMS and beyond the terminal age main sequence, as well as hints to two ridges along which a high number of radial orders seem to be excited. The first one lies close to the red edge of the $\delta$ Scuti instability strip while the second lies parallel to it but shifted towards higher temperatures. A noticeable drop in the number of excited orders can be found in between these two ridges, possibly suggesting the mechanisms behind these pulsations might differ. The `gap' remains tentative, but could be significant since there are a non-negligible number of stars in this portion of the HR diagram. Comparing the number of stars present in the gap with those in the nearby regions, it appears to be sampled consistently with the rest of the diagram. The fact that pulsating stars are numerous in the vicinity also supports the idea that this gap might not be an artifact of the sliding window method.

The uncertainties determined for the pulsation constants $Q$ are quite large, due to the large uncertainties on surface gravities in \textit{Gaia} DR3 \citep{ApsisII}. This leads to broad intervals in the amplitude spectra within which the radial fundamental and subsequent orders can be expected to lie, hence complicating the identification of excited orders of p-modes. We therefore disregard the errors for now, much like what is done in \cite{Zwintz2020}, and instead assume the number of excited orders determined from our analysis is accurate down to $\pm 1$ order. These uncertainties are important to remember when dealing with more evolved stars because as a star evolves, the radius gets larger, the density lowers, and the distance between consecutive radial orders gets smaller. 

\subsection{Metallicism}
Using the Renson spectral indexes, we computed the metallicism index defined as $\Delta = m - k$ for the 868 stars with two values. Figure \ref{Fig:MetalComp} shows the distribution of our sample according to their metallicism, as well as that of pulsating stars only. The spectral types provided in the Renson catalogue show a typical range of $\Delta$ within 2 to 12. The pulsating stars span a slightly more contrained interval. The fractional distribution of pulsating stars can be found in Fig. \ref{Fig:MetalFrac}. The errorbars were estimated using Bernoulli trials as in Equation \ref{eq:Bernoulli}:
\begin{equation}
    \sigma_f = \displaystyle\sqrt{\frac{f(1-f)}{n}}
    \label{eq:Bernoulli}
\end{equation}
$f$ is the observed fraction and $n$ is the number of stars in the bin. A downward trend can be seen for the incidence of pulsation as a function of metallicism, much like what was observed in \cite{Smalley2017}. In comparison, we note a higher incidence of pulsation at $\Delta$ indexes of 6 and less, but a lack of stars below $\Delta = 3$. The uncertainties are quite high for the bins where few stars are found, so the high fraction at $\Delta = 13$ is likely an anomaly. 

\begin{figure}
    \centering
    \includegraphics[scale=0.2]{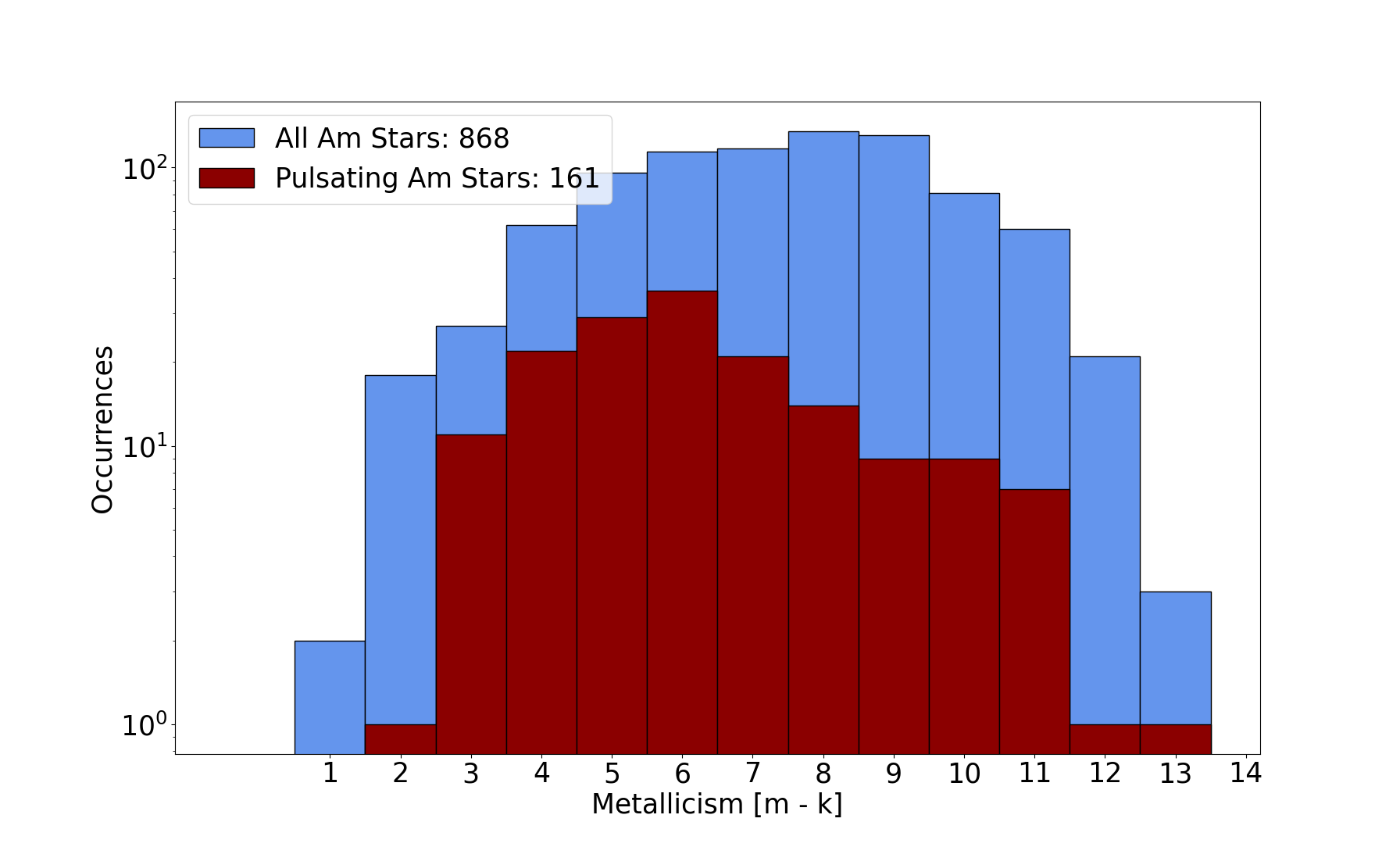}
    \caption{Distribution of metallicism in the studied population of Am/Fm stars, based on the subset of stars for which spectral indexes could be retrieved from the Renson catalogue. The entire subset is shown in blue. In red, the distribution of pulsating stars is provided.}
    \label{Fig:MetalComp}%
\end{figure}

\begin{figure}
    \centering
    \includegraphics[scale=0.2]{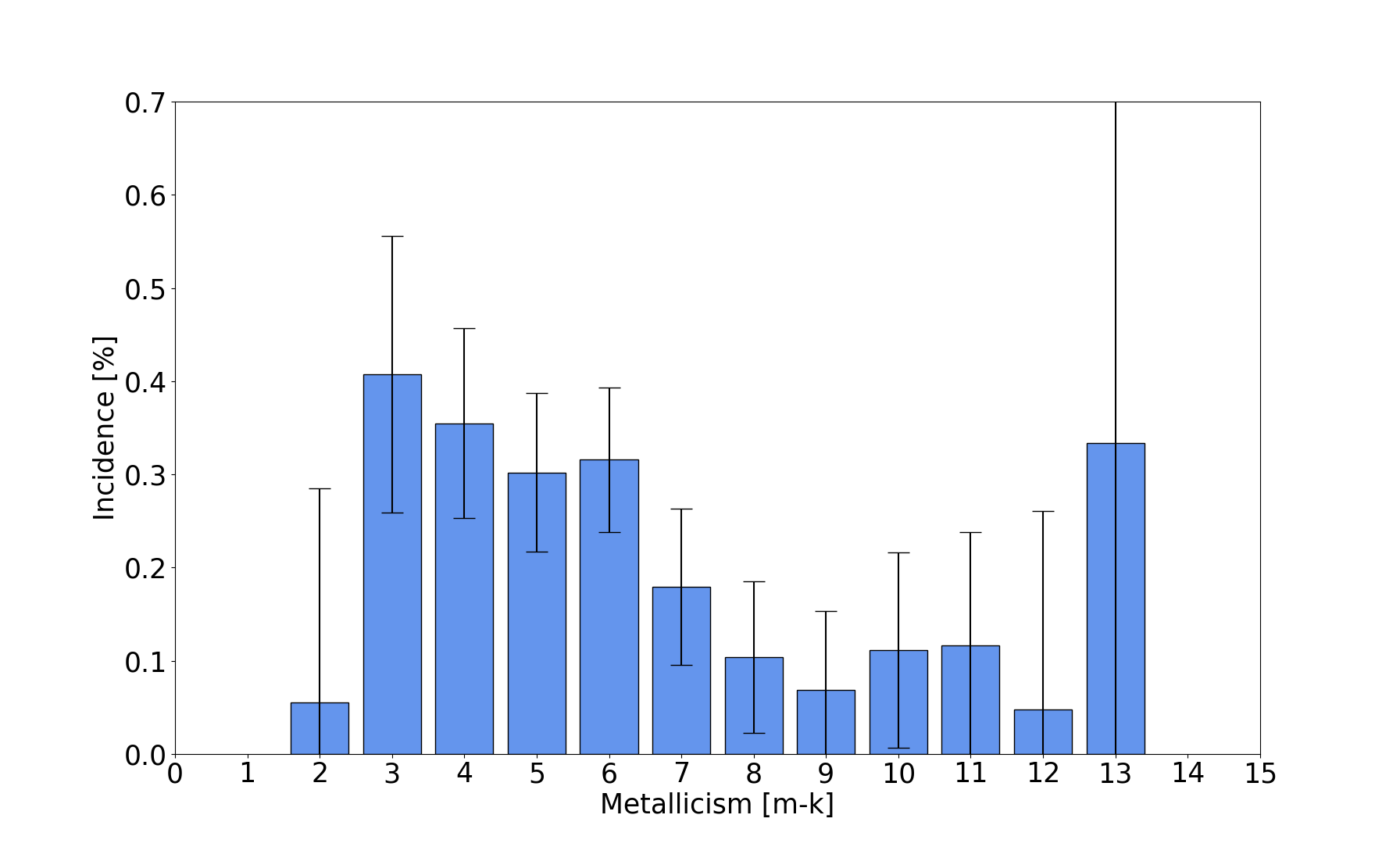}
    \caption{Fractional distribution of pulsating stars as a function of metallicism. It is clear that pulsation incidence is higher in stars with low metallicism, i.e. stars with less pronounced chemical peculiarities.}
    \label{Fig:MetalFrac}%
\end{figure}

%--------------------------------------------------------------------
\section{Discussion}
\subsection{Sample representativeness}
In this work we have analysed data from a large number of targets with the hopes of characterising the pulsation in Am/Fm stars. However, it is important to consider whether our sample is representative of the entire Am/Fm population. In the selection of our Am/Fm stars, the TESS and \textit{Gaia} detection limits are not entirely compatible, as TESS can detect brighter stars that \textit{Gaia} will not observe. In this way we can expect brighter targets to not have made the selection cut, although it is not expected to have any statistical influence on the results from our analysis. Some additional comments on the Renson catalogue are also important to take into consideration. Firstly, the spectral classification of Am/Fm stars in the Renson catalogue is in some cases doubtful, and this is taken into consideration in the catalogue by marking these with a "?" \citep{Renson}. These stars only constitute a small fraction of the entire sample and experience from WASP studies shows that their exclusion would lead to similar results. Comparing our sample with a subset of new classifications being carried out at the moment seems to be confirming that our pulsating targets are Am/Fm stars (Posilek et al. in prep) and that our sample is representative. Secondly, the presence of $\rho$ Puppis stars in the sample also affects the statistics. However, we included these in the general analysis, distinguishing between classes where necessary to avoid biased conclusions. 

\begin{figure*}
    \centering
    \sidecaption
    \includegraphics[width=12cm]{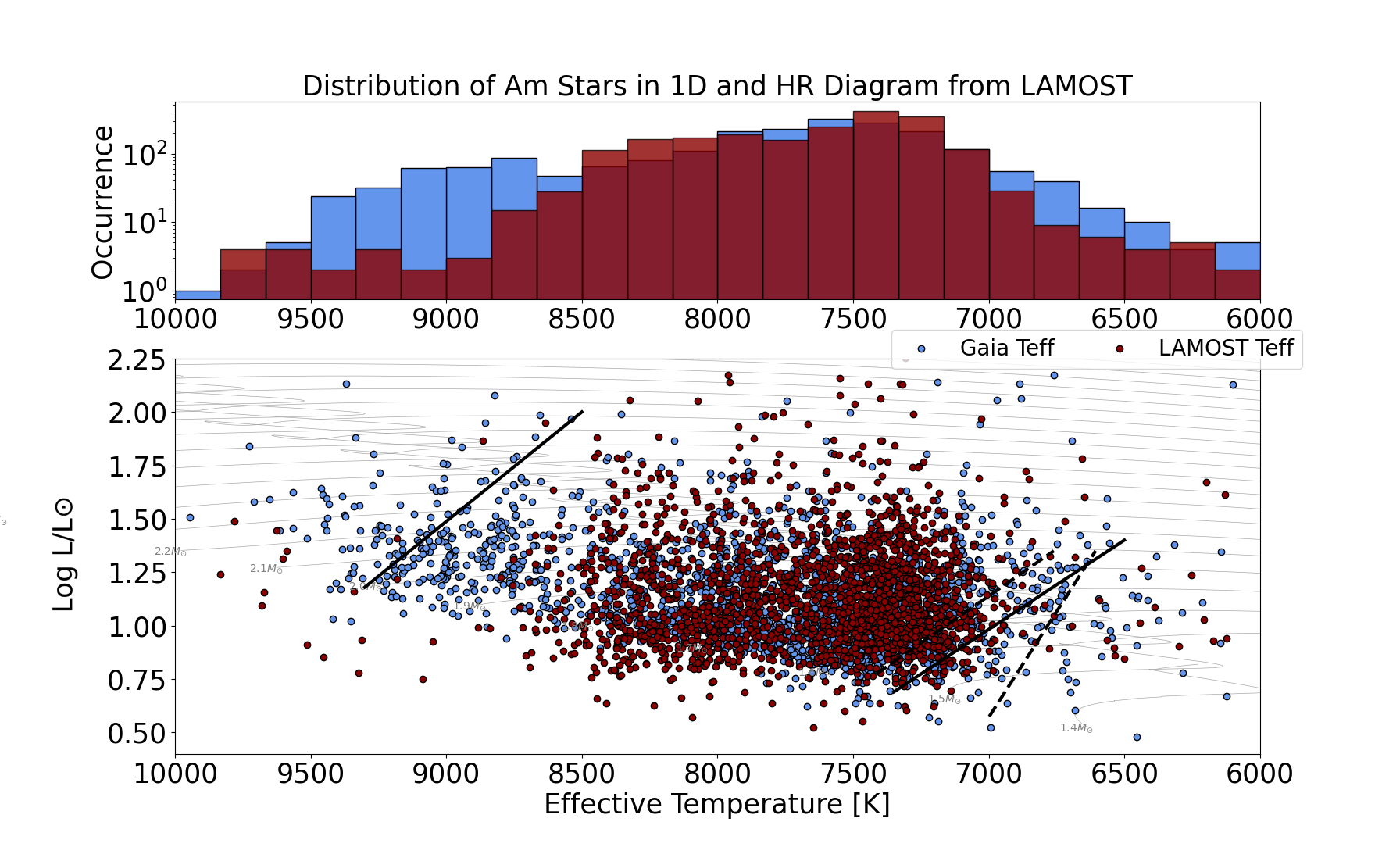}
    \caption{HR diagram containing a LAMOST sample of Am/Fm stars. Instability strips and evolutionary tracks are the same as in Fig. \ref{Fig:HRD}. In red, the sample is shown using the LAMOST effective temperatures. In blue, \textit{Gaia} DR3 GSP-Phot values are used. The latter seems to have a tendency to overestimate $T_{\rm{eff}}$, although the distributions mostly agree within the instability strips.}
    \label{Fig:LAMOST}%
\end{figure*}

We present in this paper a study of pulsation in 1276 Am/Fm stars, involving many targets for which very few or no studies have been made before. The analysis of our sample compiles known stellar parameters and provides new insight in many of these targets. Future research into the nature of Am/Fm stars will benefit greatly from spectroscopic observations. It will allow for determining stellar parameters with greater precision and detailed modelling of stars to better understand stellar oscillations and the mechanisms behind these. Work in this direction has already begun, with spectroscopy being carried out for a large number of Am/Fm stars by the Large Sky Area Multi-Object Fibre Spectroscopic Telescope (LAMOST) \citep{LAMOST_Am}. Observations have so far largely been conducted in the Galactic anti-centre \citep{LAMOST_Luo}, within longitudes 210\textdegree\ $\geq l \geq$ 150\textdegree\ and latitudes $|b| \leq 30$\textdegree\ \citep{LAMOST_Yuan}, which therefore leads to little overlap with our sample. Nevertheless, we placed the LAMOST targets in the HR diagram using their derived effective temperatures and also obtained \textit{Gaia} DR3 values to compare the sample distribution and look for potential biases (see Fig. \ref{Fig:LAMOST}). Luminosities are obtained from \textit{Gaia} in both cases. We note that GSP-Phot does indeed seem to overestimate the temperature in some cases, but that there is overall agreement between both distributions. LAMOST provides a large sample of stars with derived stellar parameters, already having proved its value for studying stellar pulsations \citep{Smalley2017}, and will be a valuable database to complement the analysis of Am/Fm population characteristics in the future. 

\subsection{Frequency analysis}
Firstly, some aspects of the variability classification should be discussed. Stars classified as likely modulated should be investigated more thoroughly if they are to be used for any following analysis. The morphology of stellar spot modulation is similar to that of ellipsoidal variability due to binarity. We could confirm the binary nature of some stars through literature and classify these correctly, but in principle all should be investigated by folding the light curves with the main period found in the amplitude spectrum. For binary systems, this should result in a perfectly periodic signal because it is governed by the orbital motion, whereas short timescale variability is likely to be observed in the case of stellar spots. Stellar spots on Am/Fm stars provide evidence of magnetic fields, however, these could also stem from lower-mass companions. 

We confined our frequency analysis to p-modes because many amplitude spectra were affected by insufficient data. This lack of data influences signals at low frequencies, increasing the risk of misclassification of $\gamma$ Doradus pulsators. This is because g-modes occupy the same frequency region and, if unresolved, can exhibit similar trends to those caused by flux modulation due to stellar spots. Because most $\gamma$ Doradus Am/Fm stars only had 1--3 sectors of data, we were not able to resolve frequencies well enough to determine radial orders for g-modes. Analysing these oscillations will be valuable in the future, when additional data are available, as these probe the near-core region, thus providing further insight in the structure of Am/Fm stars.

We focused on the ensemble analysis of stellar oscillations rather than boutique modelling of individual targets, which is impractical for such a sample size. For performing the latter, thorough validation of the frequency extraction and more detailed mode identification will be necessary. This should provide valuable insight into the dynamics of these stars and help us relate these to stellar oscillations, but require a very methodical approach to the frequency analysis, since ambiguous peaks and observational uncertainties can significantly impact the outcome \citep{Bowman2021}. 

Comparison of the observational data regarding pulsation in Am/Fm stars with theoretical models for turbulent pressure driving oscillations in Am/Fm stars suggests that more parameters need to be explored. The grid published in \cite{Antoci2019} is not dense enough for detailed analyses and does not extend down to the ZAMS, hence not covering the region of the HR diagram where most observed stars are located. High radial order excitation around 7500~K and 1.5--1.6 $L_{\odot}$ is also found in Fig. \ref{Fig:OrdersHeatmap}. However, the decrease in the number of excited radial orders between the ridges in Fig. \ref{Fig:OrdersHeatmap} is not resolved, and the excitation near the red edge is different. This assumes that no major selection bias has caused this difference in distribution of Am/Fm stars and that the targets have reasonable values for stellar parameters. It is also important to note that the heat map needs to balance many important aspects in order to show meaningful information. One important challenge is that the stars are not distributed evenly, but the radial excitation values need to be extrapolated to a uniform grid to be visualised as in Fig. \ref{Fig:OrdersHeatmap}. Choosing too small a window to project the data onto this grid results in spurious variations and does not allow for a meaningful interpretation. Instead, a sensible compromise has to be made to highlight the global trends within our sample whilst remaining sensible to rapid changes in radial order excitation within a smaller region of the HR diagram. Moreover, the grid is smoothed with a Gaussian kernel, which leads to possible shifts between the colour maxima and minima compared to the distribution of the data points. This can especially be seen in the region around $T_{\rm{eff}} \approx 7000$K and $L \approx 1.3 L_\odot$. The result presented in Fig. \ref{Fig:OrdersHeatmap} shows the best outcome of this optimisation process. Despite some limitations, there is an overall trend in the data, which is efficiently highlighted using this technique and was also identifiable in previous iterations. Improvements can be made to the method in the future to strengthen the reliability of the heat map. In parallel, verification of the existence of an excitation gap between the two main ridges will be fundamental in future work  which is in preparation (Antoci et al. in prep.), as it may suggest that different dominating pulsation mechanisms ($\kappa$-mechanism versus turbulent pressure) could be at play in Am/Fm stars. 

Precisely determining the number of excited radial orders can be challenging when working with a large sample of stars in different evolutionary stages. The predicted frequencies for each order using the \textit{Q} constants do not vary considerably with effective temperature and surface gravity within each star in the sample. However, as pulsating stars evolve from the ZAMS, the p-modes shift towards lower frequencies \citep{Pamy2000}, resulting in a higher density of radial orders than what we observe. Consequently, excited radial orders may have been missed for a number of stars in our sample. This suggests that we can expect a higher maximal number of radial orders to be truly excited particularly in evolved Am/Fm stars, similarly to what was derived in \cite{Antoci2019}.

As He depletion does indeed hinder the $\kappa$ mechanism, the amount of depletion must be an essential parameter, as various levels of depletion may affect the range of excited modes differently. In addition, the depth to which chemical peculiarities extend below the surface may also impact the pulsation characteristics. Furthermore, magnetic fields detected in a few Am/Fm stars \citep{Petit2011, Blazere2016, Neiner2017} may also play a decisive role in the excitation of pulsations. In-depth modelling of Am/Fm stars will be fundamental to investigating said properties, ultimately helping us develop our understanding of stellar oscillations and stellar interiors in general. More detailed calculations of the excitation in Am/Fm stars are in progress and will be subject to a future paper (Antoci et al. in prep). In the context of high-radial order g-mode pulsations ($\gamma$ Dor and hybrid stars), it is essential to conduct similar studies to understand how, and if at all, diffusive effects observed in Am/Fm stars affect g-mode excitation.

%-----------------------------------------------------------------
\section{Conclusions}
We analysed data from TESS and \textit{Gaia} for 1276 Am/Fm stars taken from the Renson catalogue, determining their variability from their time series and amplitude spectra. We identified binary systems, stars suspected of light curve modulation due to stellar spots, and extracted frequencies for stars suspected to pulsate. The latter was done for 25\% of the sample. Among this significant population of Am/Fm and $\rho$ Puppis stars, 90\% host p-modes, with 114 hybrids also being suspected of g-modes oscillations. We find 32 pure $\gamma$ Doradus pulsators. Placing these stars in the HR diagram, it seems that most pulsating Am/Fm stars are located near the main sequence and close to the red edge of the instability strip. We used the extracted frequencies to determine the number of excited radial orders for stars pulsating in p-modes and show the distribution of excitation in the HR diagram. Observational data suggest that excitation of a high number of radial orders can be found mainly at the ZAMS and along two ridges extending to later stages of evolution. This last aspect of the Am/Fm pulsation behaviour, as well as implications for the mechanisms at work in Am/Fm stars, will need to be studied in further detail in future work. Our sample provides the community with a number of pulsating Am/Fm stars to study further, and we believe detailed modelling of a handful of stars can provide valuable insight into stellar oscillations and chemical peculiarities in general, as well as help shed light onto the mechanisms driving pulsation in Am/Fm stars. Stellar catalogs and summary tables of the studied sample are available online.

\begin{acknowledgements}
      This work has made use of data from the European Space Agency (ESA) mission
{\it Gaia} (\url{https://www.cosmos.esa.int/gaia}), processed by the {\it Gaia}
Data Processing and Analysis Consortium (DPAC,
\url{https://www.cosmos.esa.int/web/gaia/dpac/consortium}). Funding for the DPAC
has been provided by national institutions, in particular the institutions
participating in the {\it Gaia} Multilateral Agreement. Co-funded by the European Union (ERC, MAGNIFY, Project 101126182 ). Views and opinions expressed are however those of the author(s) only and do not necessarily reflect those of the European Union or the European Research Council. Neither the European Union nor the granting authority can be held responsible for them.

Reproduced with permission from Astronomy \& Astrophysics, © ESO
\end{acknowledgements}

\bibliographystyle{aa} 
\bibliography{citations}

\begingroup
\setlength{\tabcolsep}{4pt}
\begin{sidewaystable*}
\appendix
\section{Additional tables}
\caption{Excerpt of the summary table for our sample of 1276 Am/Fm stars.}\label{tab:AmStars}
\centering
\begin{tabular}{lllcccccccccccc} 
    \hline
    \noalign{\smallskip}
    TIC & \textit{Gaia} ID & Alternative ID & $T_{\rm{eff}}$ [K] & $\log g$ [$g_{\odot}$] & $L$ [$L_{\odot}$] & $\sigma L$ & V mag & Variability & g-modes [d$^{-1}$] & p-modes [d$^{-1}$] & $N_{\rm{radial}}$ & Binarity & $N_{\rm{sectors}}$ & $\rho$\,Pup. \\
    \noalign{\smallskip}
    \hline
    \noalign{\smallskip}
    ... & ... & ... & ... & ... & ... & ... & ... & ... & ... & ... & ... & ... & ...  & ... \\
    \noalign{\smallskip}
    151681127 & 5390443239364017024 & HD 96645  & 9062  & 4.3 & 18.2 & 1.2 & 9.26  & EB & -              & -                & - & EB & 1 & - \\
    \noalign{\smallskip}
    151769040 & 5389362247635466624 & HD 97160  & 7110  & 4.0 & 9.2  & 0.6 & 8.49  & HY & {[}1.0, 5.0{]} & {[}5.1, 348.3{]} & 3 & -  & 3 & - \\
    \noalign{\smallskip}
    152475194 & 4838691708790468352 & HD 26507  & 7471  & 4.2 & 6.8  & 0.5 & 8.99  & OV & -              & -                & - & -  & 2 & - \\
    \noalign{\smallskip}
    152864226 & 6553048898788178304 & HD 217417 & 7125  & 3.9 & 8.0  & 0.5 & 8.11  & HY & {[}1.0, 5.0{]} & {[}5.1, 348.3{]} & 7 & -  & 2 & - \\
    \noalign{\smallskip}
    153798808 & 4839725112282144640 & HD 28538  & 7164  & 4.0 & 9.5  & 0.6 & 8.44  & -  & -              & -                & - & -  & 4 & - \\
    \noalign{\smallskip}
    153860833 & 1124869634384221824 & HD 71973  & 7288  & 3.9 & 13.7 & 0.9 & 6.27  & OV & -              & -                & - & -  & 4 & - \\
    \noalign{\smallskip}
    154568099 & 1691969621428898688 & HD 110985 & 7829  & 3.6 & 45.6 & 3.1 & 8.53  & DS & -              & {[}5.1, 45.7{]}  & 4 & -  & 5 & - \\
    \noalign{\smallskip}
    155785420 & 870514799169190016  & HD 56152  & 8934  & 3.8 & 56.2 & 3.8 & 7.34  & OV & -              & -                & - & -  & 2 & - \\
    \noalign{\smallskip}
    156576373 & 3342526141695647744 & HD 41076  & 10230 & 3.8 & 80.1 & 5.4 & 6.075 & -  & -              & -                & - & -  & 1 & - \\
    \noalign{\smallskip}
    157016177 & 3999857073730936448 & * 2 Com A & 7474  & 3.6 & 29.8 & 2.0 & 6.15  & DS & -              & {[}5.1, 45.7{]}  & 6 & -  & 2 & - \\
    \noalign{\smallskip}
    ... & ... & ... & ... & ... & ... & ... & ... & ... & ... & ... & ... & ... & ... & ... \\
    \noalign{\smallskip}
    \hline
\end{tabular}
\tablefoot{Each entry describes one star. We chose to include three identifiers to facilitate cross-identification: the TIC identifier, the \textit{Gaia} DR3 identifier and an alternative identifier, most commonly the HD ID. The next 5 columns describe key stellar parameters: effective temperature, surface gravity, luminosity and associated uncertainty, the apparent magnitude in V band. The uncertainties for effective temperature and surface gravity are not included since these are fixed to 110 K and 0.2 dex (see Section \ref{sec:Gaia}). Next, 4 columns summarise the results obtained from our frequency analysis: the variability class if pulsations were detected, the range of frequencies excited for g and p modes if applicable and the number of excited radial orders identified. The variability classes are DS for $\delta$ Scuti, GD for $\gamma$ Doradus, HY for hybrids, EB for eclipsing binaries, and OV for other variability. Lastly, information is provided on the potential binary nature of the system, the number of TESS sectors used for the analysis and whether the star is a $\rho$\,Puppis. The full table is available at the CDS.}
\end{sidewaystable*}
\end{document}